\documentclass{aa}
\pdfoutput=1

\usepackage{graphicx}
\usepackage{subfig}
\usepackage[varg]{txfonts}
\usepackage{lscape}
\usepackage{color}
\usepackage{longtable}

\def\m2s2{\hbox{\,m$^{2}$\,s$^{-2}$}} 

\begin{document}

   \title{The HADES RV Programme with HARPS-N at TNG\thanks{Based on: observations made with the Italian \textit{Telescopio Nazionale Galileo} (TNG), operated on the island of La Palma by the INAF - \textit{Fundaci\'on Galileo Galilei} at the \textit{Roche de Los Muchachos} Observatory of the \textit{Instituto de Astrof\'isica de Canarias} (IAC); photometric observations made with the APACHE array located at the Astronomical Observatory of the Aosta Valley; photometric observations made with the robotic telescope APT2 (within the EXORAP programme) located at Serra La Nave on Mt. Etna.}}
   \subtitle{XI. GJ 685\,b: a warm super-Earth around an active M dwarf}
   \titlerunning{GJ 685}

   \author{M. Pinamonti\inst{\ref{inst3}}
          \and
          A. Sozzetti\inst{\ref{inst3}}
          \and
          P. Giacobbe\inst{\ref{inst3}}
          \and
          M. Damasso\inst{\ref{inst3}}
          \and
          G. Scandariato\inst{\ref{inst7}}
          \and
          M. Perger\inst{\ref{inst13},\ref{inst15}}
          \and
          J. I. Gonz\'alez Hern\'andez\inst{\ref{inst11},\ref{inst12}}
          \and
          A. F. Lanza\inst{\ref{inst7}}
          \and
          J. Maldonado\inst{\ref{inst6}}
          \and
          G. Micela\inst{\ref{inst6}}
          \and
          A. Su\'arez Mascare\~no\inst{\ref{inst11},\ref{inst14}}
          \and
          B. Toledo-Padr\'on\inst{\ref{inst11},\ref{inst12}}
          \and
          L. Affer\inst{\ref{inst6}}
          \and
          S. Benatti\inst{\ref{inst5}}
          \and
          A. Bignamini\inst{\ref{inst2}}
          \and
          A. S. Bonomo\inst{\ref{inst3}}
          \and
          R. Claudi\inst{\ref{inst5}}
          \and
          R. Cosentino\inst{\ref{inst7},\ref{inst8}}
          \and
          S. Desidera\inst{\ref{inst5}}
          \and
          A. Maggio\inst{\ref{inst6}}
          \and
          A. Martinez Fiorenzano\inst{\ref{inst8}}
          \and
          I. Pagano\inst{\ref{inst7}}
          \and
          G. Piotto\inst{\ref{inst5},\ref{inst4}}
          \and
          M. Rainer\inst{\ref{inst16}}
          \and
          R. Rebolo\inst{\ref{inst11},\ref{inst12}}
          \and
          I. Ribas\inst{\ref{inst13},\ref{inst15}}
          }

   \institute{INAF - Osservatorio Astrofisico di Torino, Via Osservatorio 20, I-10025 Pino Torinese, Italy\\
              \email{m.pinamonti.astro@gmail.com}\label{inst3}
         \and
             INAF - Osservatorio Astrofisico di Catania, Via S. Sofia 78, I-95123 Catania, Italy\label{inst7}
         \and
             Institut de Ci\`encies de l'Espai (ICE, CSIC), Campus UAB, C/ de Can Magrans s/n, E-08193 Cerdanyola del Vall\`es, Spain\label{inst13}
         \and
             Institut d'Estudis Espacials de Catalunya (IEEC), C/ Gran Capit\`a 2-4, E-08034 Barcelona, Spain\label{inst15}
         \and
             Instituto de Astrof\'isica de Canarias (IAC), E-38205 La Laguna, Tenerife, Spain\label{inst11}
         \and
             Universidad de La Laguna, Dpto. Astrof\'isica, E-38206 La Laguna, Tenerife, Spain\label{inst12}
         \and
             INAF - Osservatorio Astronomico di Palermo, piazza del Parlamento 1, I-90134 Palermo, Italy\label{inst6}
         \and
             Observatoire Astronomique de l'Universit\'e de Gen\'eve, 1290 Versoix, Switzerland\label{inst14}
         \and
             INAF - Osservatorio Astronomico di Padova, vicolo dell'Osservatorio 5, I-35122 Padova, Italy\label{inst5}
         \and
             INAF - Osservatorio Astronomico di Trieste, via G. B. Tiepolo 11, I-34143 Trieste, Italy\label{inst2}
         \and
             Fundaci\'on Galileo Galilei - INAF, Ramble Jos\'e Ana Fernandez P\'erez 7, E-38712 Bre\~na Baja, TF, Spain\label{inst8}
         \and
             Dipartimento di Fisica e Astronomia, Universit\`a di Padova, via Marzolo 8, I-35131 Padova, Italy\label{inst4}
         \and
             INAF - Osservatorio Astrofisico di Arcetri, Largo E. Fermi 5, I-50125 Firenze, Italy\label{inst16}
         }

   \date{Received <date> /
      Accepted <date>}

  \abstract
   {Small rocky planets seem to be very abundant around low-mass M-type stars. Their actual planetary population is however not yet precisely understood. Currently several surveys aim to expand the statistics with intensive detection campaigns, both photometric and spectroscopic.}
   {The HADES program aims to improve the current statistics through the in-depth analysis of accurate radial velocity monitoring in a narrow range of
spectral sub-types, with the precision needed to detect small planets with a few Earth masses. }
   {We analyse  106 spectroscopic HARPS-N observations of the active M0-type star GJ 685 taken over the past five years. We combine these data with photometric measurements from different observatories to accurately model the stellar rotation and disentangle its signals from genuine Doppler planetary signals in the RV data. We run an MCMC analysis on the RV and activity indexes time series to model the planetary and stellar signals present in the data, applying Gaussian Process regression technique to deal with the stellar activity signals.}
   {We identify three periodic signals in the RV time series, with periods of 9, 24, and 18 d. Combining the analyses of the photometry of the star with the activity indexes derived from the HARPS-N spectra, we identify the 18 d and 9 d signals as activity-related, corresponding to the stellar rotation period and its first harmonic respectively. The 24 d signals shows no relations with any activity proxy, so we identify it as a genuine planetary signal. We find the best-fit model describing the Doppler signal of the newly-found planet, GJ 685\,b, corresponding to an orbital period $P_b = 24.160^{+0.061}_{-0.047}$ d and a minimum mass $M_P \sin i = 9.0^{+1.7}_{-1.8}$ M$_\oplus$. We also study a sample of 70 RV-detected M-dwarf planets, and present new statistical evidence of a difference in mass distribution between the populations of single- and multi-planet systems, which can shed new light on the formation mechanisms of low-mass planets around late-type stars.}
   {}
   

   \maketitle
%

\section{Introduction}

Most of the early surveys hunting for exoplanets, which employed the radial velocity method, directed their efforts towards dwarf stars of spectral type G or K, usually in a range around the mass of the Sun \citep[e.g.][]{quelozetal2001,valentifischer2005,tamuzetal2008}. Instead in recent years M dwarfs have become the most promising targets for the hunt for low-mass, rocky planets \citep[e.g.][]{bonfils13,dreschar2013,sozzettietal13,astudillodefruetal2017}, due to their more advantageous mass and radius ratios compared to solar-type stars. Moreover, with the availability of high-precision spectrographs mounted on 4 m-class telescopes, led by HARPS at La Silla \citep[Northern High Accuracy Radial velocity Planet Searcher,][]{mayoretal2003} and recently backed up by its younger twin HARPS-N at TNG \citep{cosentinoetal2012}, it was possible to reach 1 m$/$s precision allowing the detection of small Earth-like rocky planets \citep[e.g.][]{anglada16b,astudillodefruetal2017b}.

It is becoming clear that giant gas planets are less frequent around low-mass than around Solar-type stars, as expected from theoretical studies \citep[e.g.][]{laughlinetal2004,mordasinietal2009}, while low-mass rocky planets are showing to be much more common than around solar-type stars, both in Radial Velocities (RVs) \citep[e.g.][and references therein]{tuomi14} and transits \citep[e.g.][and references therein]{gaidosetal2016} observations.
Nevertheless, the complete characterization of this abundant population of rocky planets around M dwarfs is hindered by the strong effects of the stars' photospheric and magnetic activity, which can produce RV signals as large as tens of m s$^{-1}$. This can result in stellar signals being mistaken for planetary signals or otherwise uncertain results for systems around active stars \citep[e.g.][]{bonfilsetal2007,baluev2013b,robertsonetal2014},  and also around quieter targets, which can still present periodic signals of unclear nature \citep[e.g.][]{robertsonetal2015b,anglada16a}.

The Harps-n red Dwarf Exoplanet Survey (HADES) programme is a collaboration between the Italian Global Architecture of Planetary Systems \citep[GAPS,][]{covinoetal2013,desideraetal2013,porettietal2016} Consortium\footnote{\url{http://www.oact.inaf.it/exoit/EXO-IT/Projects/Entries/2011/12/27_GAPS.html}}, the Institut de Ci\`encies de l'Espai de Catalunya (ICE), and the Instituto de Astrof\'isica de Canarias (IAC). The aim of the survey is to characterize the exoplanetary systems around a well-defined sample of M dwarfs, with spectral type between dM0 and dM3. High-precision RVs of the sample have been collected over the course of five years with the HARPS-N@TNG spectrograph. Several planets have been discovered as part of the survey \citep[e.g.][]{afferetal2016,suarezmascarenoetal2017,pergeretal2017b,pinamontietal2018}, and also some more general studies have already been performed on the samples, both studying the stellar properties \citep{maldonadoetal2017,scandariatoetal2017,suarezmascarenoetal2018,gonzalezalvarezetal2019} and the preliminary planetary population statistics \citep{pergeretal2017}.

In this work we present the search for planetary companions around the M dwarf GJ 685, based on high precision spectroscopic observations carried out with HARPS-N as part of the HADES programme. We also take advantage of ancillary photometric observations of the target to better constrain the stellar activity signal in the RVs.

In Sect. \ref{rv_time_series} we describe the Doppler measurements of GJ 685 collected for this analysis, and in Sect. \ref{stellar_prop} we briefly discuss the physical properties of the host star. The independent analyses of two photometric datasets are presented in Sect. \ref{photo_analysis}. We describe our periodogram analyses of the RV data and stellar activity indexes in Sect. \ref{periodogram_sec}. We proceed to find the best-fit parameters for the models describing the activity and RV time series via an MCMC analysis in Sect. \ref{mcmc_model}. Finally, we summarize and discuss our findings in the more general context of the current population of M-dwarf RV-detected planetary systems  in Sect. \ref{paperm76_conclusions}.


\section{Spectroscopic observations}
\label{rv_time_series}

As part of the HADES RV programme, GJ 685 has been observed from BJD $=2456439.6$  (27th May 2013) to BJD $= 2458044.4$ (17th October 2017), with the HARPS-N spectrograph, connected by fibers to the Nasmyth B focus through a Front End Unit of the 3.58m Telescopio Nazionale Galileo (TNG) in La Palma, Spain. HARPS-N is a fiber-fed, cross-dispersed echelle spectrograph with a spectral resolution of $115\;000$, covering a wavelength range from 3830 to 6900 $\AA$. We observed with fixed integration times of 900 s to obtain data of sufficient signal-to-noise ratio (SNR > 20) and to average out short-term periodic oscillations of the star, such as p-modes \citep{dumusqueetal2011}.

\begin{figure}
   \centering
   \includegraphics[width=.45\textwidth]{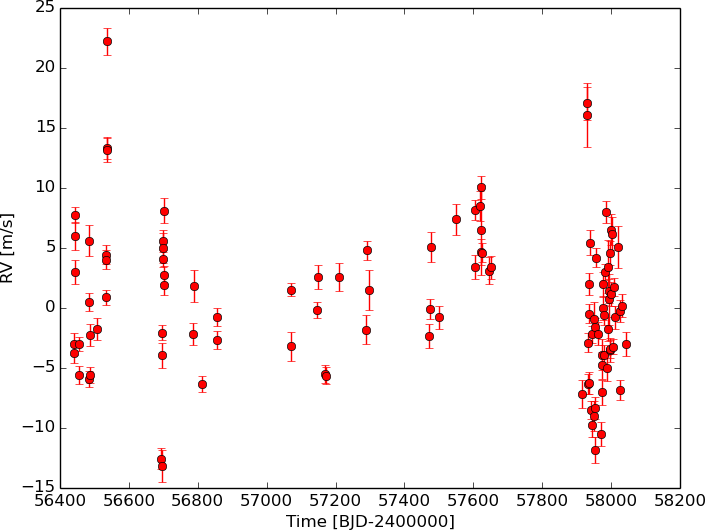}
      \caption{HARPS-N RV time series of GJ 685.}
         \label{time_series}
\end{figure}

The total number of data points acquired was 106 over a time span of $1605$ days. The time series is shown in Figure \ref{time_series}. The observations were gathered without the simultaneous Th-Ar calibration, which could contaminate the Ca~{\sc ii}  H \& K lines due to the long exposure times and the relative faintness of M-dwarf targets in the blue part of the spectra. Moreover, this lines are crucial in the analysis of stellar activity \citep[e.g.][]{giampapaetal1989,forveilleetal2009}, which is particularly important for active late-type stars. Thus a precise acquisition of the Ca~{\sc ii}  H \& K lines was preferred over a better correction of possible instrumental drifts.
\citet{pergeretal2017} used other GAPS target spectra, gathered by the Italian team during the same nights using the simultaneous Th-Ar calibration, to quantify this effect, and found a mean inter-night instrumental drift of the order of $1$ m s$^{-1}$ over the whole HADES sample.
Nevertheless, for some of the brightest targets of the sample, some spectra could be safely taken with the simultaneous drift calibrations without problems for the Ca~{\sc ii}  H \& K observations \citep{pergeretal2017b}, which helped monitor the instrumental drift over the single target: for GJ 685, 4 spectra were collected with the simultaneous Th-Ar calibrations, measuring a mean inter-night instrumental drift of $\sim 0.5$ m s$^{-1}$. Any residual drift in the time series has been taken into account in our final model as discussed in Sect. \ref{emcee_analysis}.
 
The data reduction and the RV extraction were performed using the HARPS-N Data Reduction Software \citep[DRS,][]{lovispepe2007} and the TERRA pipeline \citep[Template-Enhanced Radial velocity Re-analysis Application,][]{anglada-escudebutler2012} respectively, which is considered to be more accurate when applied to M-dwarfs, with respect to the DRS. For a more thorough discussion of the DRS and TERRA performances on the HADES targets see \citet{pergeretal2017}. The mean internal error of the TERRA data is $1.02$ m s $^{-1}$, with a few low-SNR data with $\sigma_\text{RV} > 2.0$ m s$^{-1}$. With an rms of $6.16$ m s$^{-1}$, GJ 685 is one of the HADES targets with the largest RV dispersion. The TERRA pipeline also corrected the RV data for the perspective acceleration of GJ 685, ${dv_r / dt = 0.11}$ m s$^{-1}$ yr$^{-1}$.

\section{Stellar properties of GJ 685}
\label{stellar_prop}

The star GJ 685 is a high proper motion nearby ($\pi = 69.825 \pm 0.039$ mas) M0-type dwarf star. We used the stellar parameters published by \citet{maldonadoetal2017}, which were calculated applying the empirical relations by \citet{maldonadoetal2015} on the same HARPS-N spectra from which we derived the RV time series. The parallax and proper motions of the star were taken from the Gaia Data Release 2 \citep{gaiaetal2018}. All the stellar parameters of GJ 685 are listed in Table \ref{star_par}.

\citet{suarezmascarenoetal2018} studied the presence of signatures of magnetic cycles and rotation on the stars of the HADES sample, measuring the rotation periods and $\log R'_\text{HK}$ for several stars of the sample\footnote{\citet{suarezmascarenoetal2018} extended the definition of $\log R'_\text{HK}$ for application on M-dwarfs spectra, following a procedure very similar to the one used by \citet{astudillodefruetal2017c}.}.
For GJ 685, \citet{suarezmascarenoetal2018} derived, from the variability in the  \textit{S}-index and H$\alpha$ activity indexes and RV time series, a rotation period of $16.3 \pm 4.2$ d finding no evidence for the presence of a magnetic cycle. The rotation period value is listed in Table \ref{star_par} along with the measured value of $\log R'_\text{HK}$.

\begin{table}
\caption{Stellar parameters for the target GJ 685}
\small
\label{star_par}
\centering
\begin{tabular}{lc}
\hline\hline
Parameter & GJ 685\\
\hline
   Spectral type & M0.5 \tablefootmark{a} \\
   $T_{\text{eff}}$ $[$K$]$ & $3816 \pm 69$ \tablefootmark{a} \\
   $[$Fe$/$H$]$ $[$dex$]$ & $-0.15 \pm 0.09$ \tablefootmark{a} \\
   Mass $[$M$_\odot]$ & $0.55 \pm 0.06$ \tablefootmark{a} \\
   Radius $[$R$_\odot]$ & $0.54 \pm 0.05$ \tablefootmark{a} \\
   $\log g$ $[$cgs$]$ & $4.72 \pm 0.05$ \tablefootmark{a} \\
   $\log \text{L}_*/\text{L}_\odot$ & $-1.253 \pm 0.094$ \tablefootmark{a} \\
   $v \sin i$ $[$km s$^{-1}]$ & $1.33 \pm 0.42$ \tablefootmark{a} \\
   $\log R'_\text{HK}$ & $- 4.79 \pm 0.04$ \tablefootmark{b} \\
   $P_\text{rot}$ & $16.3 \pm 4.2$ \tablefootmark{b} \\
\hline                                   
$\alpha$ (J2000) & 17$^h$:35$^m$:35.0$^s$ \tablefootmark{c} \\
$\delta$ (J2000) & +61$^\circ$:40$'$:45.6$''$ \tablefootmark{c} \\
$B-V$ $[\text{mag}]$ & 1.48 \\
$V$ $[\text{mag}]$ & 9.97  \\
$J$ $[\text{mag}]$ & 6.884 \tablefootmark{d} \\
$H$ $[\text{mag}]$ & 6.271 \tablefootmark{d} \\
$K$ $[\text{mag}]$ & 6.066 \tablefootmark{d} \\
$\pi$ $[\text{mas}]$ & $69.825 \pm 0.039$ \tablefootmark{c} \\
$\mu_\alpha$ $[\text{mas yr}^{-1}]$ & $261.895 \pm 0.055$ \tablefootmark{c} \\
$\mu_\delta$ $[\text{mas yr}^{-1}]$ & $-514.400 \pm 0.063$ \tablefootmark{c} \\
\hline
\end{tabular}
\tablefoot{\tablefoottext{a}{\citet{maldonadoetal2017}}; \tablefoottext{b}{\citet{suarezmascarenoetal2018}};  \tablefoottext{c}{\citet{gaiaetal2018}}; \tablefoottext{d}{\citet{cutrietal2003}}}
\end{table}

\section{Photometric monitoring}
\label{photo_analysis}

As most of the targets of the HADES sample, GJ 685 has been monitored photometrically by means of the APACHE \citep{sozzettietal13} and EXORAP (EXOplanetary systems Robotic APT2 Photometry) surveys. The two surveys perform regular follow-up observations of HADES M-dwarf targets to constrain the stellar rotation periods by analyzing the photometric variability. We briefly discuss the analyses of the datasets collected by the two surveys in the following sections.

\subsection{APACHE photometry}
\label{apache_photo}

GJ 685 was monitored for 64 nights between BJD $= 2456456.4$ (12th Jun 2013) and BJD $= 2456793.6$ (15th May 2014), with one of the five 40cm telescopes composing the APACHE array, located at the Astronomical Observatory of the Autonomous Region of the Aosta Valley (OAVdA, $+45.7895$ N, $+7.478$ E, 1650 m.a.s.l.). The observations were collected following the standard APACHE procedure, and the images were reduced with the standard pipeline TEEPEE by the APACHE team \citep{giacobbeetal2012}.

Since \citet{suarezmascarenoetal2018} calculated the rotation period of GJ 685 to be $P_\text{rot} = 16.3 \pm 4.2$ d, we performed Generalized Lomb Scargle periodogram analysis \citep[GLS,][]{zechkur2009} of the photometric data, looking for similar periodicities. To do so we binned the data over each night to average out the short period noise due to the high number of very close data points. This resulted in a time series of 64 data points over a time span of 337 d, with an rms of 0.012 mag.

\begin{figure}
   \centering
 \includegraphics[width=.45\textwidth]{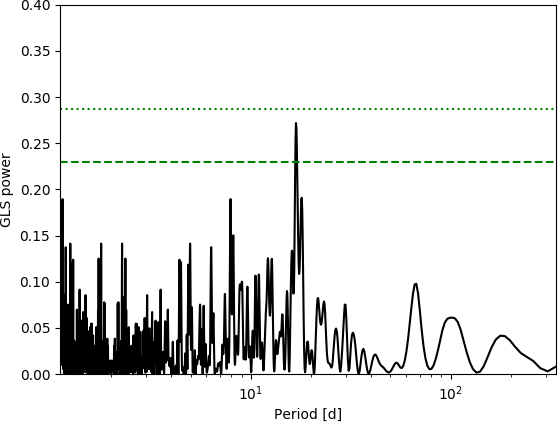} \\
 \includegraphics[width=.45\textwidth]{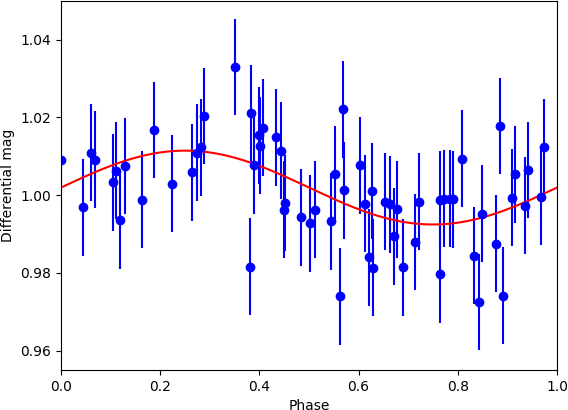}
 \caption{Upper panel: GLS periodograms of the APACHE photometric data. The dotted and dashed horizontal lines indicate the 1$\%$ and 10$\%$ FAP levels respectively. Lower panel: APACHE light curve phase-folded  over the 16.85 d period found by the GLS periodogram. The red dots represent the best-fit model at the observed epochs.}
 \label{apache_periodograms_plot}
\end{figure}

The results of the GLS analysis, which covered periods between 1 d and the time span of the time series, are shown in Fig. \ref{apache_periodograms_plot}, with the highest peak at a period of $P_\text{rot, AP} = 16.85 \pm 0.12$ d with a theoretical False Alarm Probability (FAP) of $1.9 \%$, in good agreement with the rotation period found by \citet{suarezmascarenoetal2018}. It is also worth noticing that there is no significant peak at longer periods.

\subsection{EXORAP photometry}

In the framework of the EXORAP project, we observed GJ 685 using an 80 cm f$/$8 Ritchey-Chretien robotic telescope (APT2) located at Serra la Nave on the Mt. Etna and operated by the INAF-Catania Astrophysical Observatory. We collected $\sim$200 measurements in each band between 5th May 2014 and 6th Sept 2017. Data reduction is preformed by overscan, bias, dark subtraction, and flat fielding with IRAF procedures and visually inspected to check the data quality (see \citet{afferetal2016} for details).

The scatter of the B and V photometry is slightly larger than 0.01 mag, which corresponds to the intra-night sensitivity of the survey. This suggests that there is some jitter of stellar origin in the collected data. The Pooled Variance (PV) analysis \cite[and references therein]{scandariatoetal2017} identifies a significant time scale around 20 d, but the precision of the PV technique is not sufficient to distinguish the 18 and 24 d periodicities that we will discuss in Sec. \ref{sec_rv_period}. The GLS periodogram analysis shows a low-significance peak again around 20 d, while the GP analysis converges directly to an 18 d period, excluding longer periods closer to 24 d.

The analysis of the RI photometry does not lead to any significant result, as the scatter of the data of $\sim0.01$ mag is dominated by the inter-night sensitivity of the survey. This is consistent with a scenario where the photometric scatter is dominated by cool photospheric spots, whose contrast against the unspotted photosphere is larger in the bluer bands than in R and I. This is also consistent with other similar analyses we published in other papers of the HADES series.

\section{Periodogram analyses}
\label{periodogram_sec}

We started by analysing our spectroscopic data by means of GLS periodograms, in order to identify significant periodicities in our time series, and compare the signals identified in the RV and activity indexes time series to pinpoint activity-related signals in the RV time series.

\subsection{Radial velocity periodograms}
\label{sec_rv_period}

\begin{figure}
   \centering
   \subfloat[][\emph{RV original time series}]
 {\includegraphics[width=.39\textwidth]{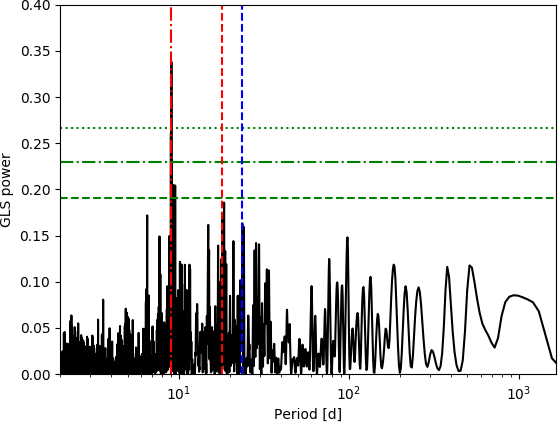}} \\
   \subfloat[][\emph{RV first residuals}]
 {\includegraphics[width=.39\textwidth]{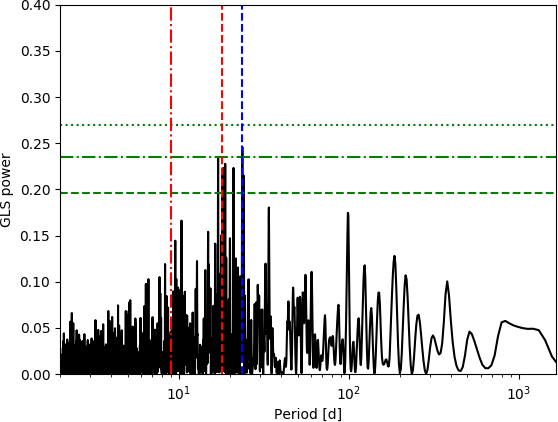}} \\
   \subfloat[][\emph{RV second residuals}]
 {\includegraphics[width=.39\textwidth]{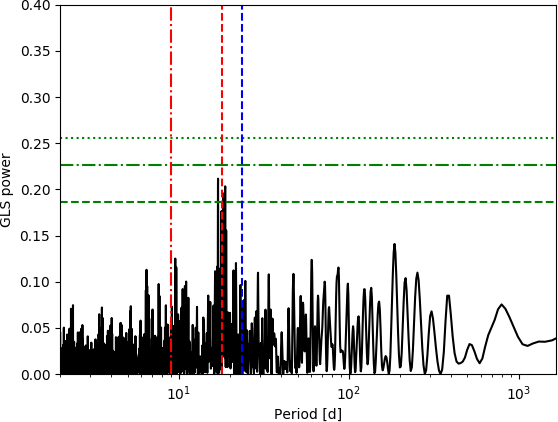}}
 \caption{GLS periodograms of the RV time series and residuals after consecutive signal fits. The red vertical lines indicate the potential rotation period (dashed) and its first harmonic (dot-dashed), while the dashed blue vertical lines marks the orbital period of the planetary candidate. The dotted, dash-dotted, and dashed horizontal lines indicate the 0.1$\%$, 1$\%$, and 10$\%$ FAP levels, respectively.}
 \label{rv_periodograms_plot}
\end{figure}

First we analysed the RV time series, identifying additional significant periodicities by pre-whitening until no significant signal was found in the GLS periodogram below the FAP $= 10\%$ level. We computed the GLS FAPs via bootstrap randomization with 10000 iterations \citep{endletal2001}. The resulting periodograms are shown in Figure \ref{rv_periodograms_plot}. The strongest peak in the first periodogram is at $P = 9$ d, which is probably due to stellar activity, being roughly half the rotation period of the star $P_\text{rot} = 16.3 \pm 4.2$ d, as derived by \citet{suarezmascarenoetal2018} and confirmed by our photometric analysis in the previous section. After subtracting this signal we see that a peak around $P =  23.66$ d rises from the periodogram, along with a clustered peak around $P = 18$ d. The latter can be related again to the activity signal of the star, proving that a simple sinusoidal fit with $P = P_{1/2} = 9$ d is not sufficient to model the influence of active regions on the RVs. We thus identify $P_\text{rot} = 18$ d as the rotation period of the star. It is worth noticing that this rotation period is longer than the one derived from the analysis of the APACHE photometry in Sect. \ref{apache_photo}. This can be explained as a consequence of two effects: first, activity signals in the photometry can differ from those present in the RV time series, depending on the nature of the active regions present on the star \citep[e.g.][]{kursteretal2003,dumusqueetal2014}; second, the rotation signal is relatively weak in the photometry, with a theoretical FAP $> 1\%$ in the APACHE analysis and not appearing clearly in the EXORAP datasets, and thus a relatively large uncertainty in the rotation period is not surprising. Moreover, it is worth noticing that a peak at $P = 18$ d is indeed present in the periodogram in Fig. \ref{apache_periodograms_plot}, even if weaker than the main peak. The $P = 23.7$ d signal, instead, does not appear to be easily related to the assumed rotation period of the star or its harmonics, and it is also relatively strong in the periodogram (FAP $= 0.47\%$). We thus suspect it to be a genuine Keplerian Doppler shift due to an orbiting planet, hereafter GJ 685\,b. The semi-amplitude of the signal at $P =  23.661 \pm 0.037$ d identified in the periodogram is $K = 3.11 \pm 0.53$ m s$^{-1}$.

For a more comprehensive view of the periodicities present in the RV data, we also studied the time series with the Bayesian Generalized Lomb-Scargle \citep[BGLS,][]{mortier2015} and FREquency DEComposer \citep[FREDEC, ][]{baluev2013fre}, to compare their different results as suggested in \citet{pinamontietal2017}. The BGLS periodogram results are very similar to those from GLS, with the 9 d peak dominating the first periodogram, and the $P = 23.7$ d signal emerging in the residuals analysis. A difference arises in the second residual analysis, compared to the bottom panel of Fig. \ref{rv_periodograms_plot}, since the periodic signals are much weaker, with a theoretical BGLS FAP (as defined in \citet{pinamontietal2017}) $> 10\%$\footnote{For the comparative analysis of the different periodogram algorithms we computed only the theoretical FAPs, since the bootstrap simulation, in particular applied to FREDEC, with be very time-consuming.}. The FREDEC analysis instead produces as best fit solution the 5-signals solution $P_1 = 18.46$ d $P_2 = 18.10$ d $P_3 = 9.66$ d $P_4 = 9.05$ d $P_5 = 8.99$ d, with a FAP$ = 1 \%$. As an alternative, if barely less significant (FAP $\simeq 1.5 \%$) solution, the algorithm proposes a 4-signals solution $P_1 = 	23.71$ d $P_2 = 17.03$ d $P_3 = 9.65$ d $P_4 = 9.04$ d, which includes also the 23.7 d signal found by the other techniques. It is worth noticing the presence of multiple periodic signals near the stellar rotation period and its first harmonic, which suggests the RV stellar activity signal to be strongly quasi-periodic.

\subsection{Stellar activity analysis}
\label{stellar_activity_periodograms}

\begin{figure}
   \centering
   \subfloat[][\emph{BIS}]
   {\includegraphics[width=.39\textwidth]{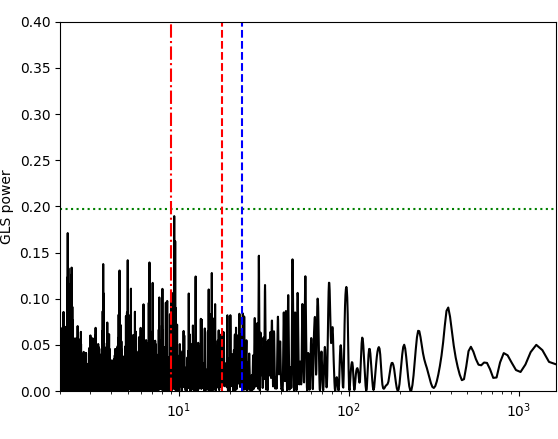}} \\
   \subfloat[][\emph{$\Delta V$}]
   {\includegraphics[width=.39\textwidth]{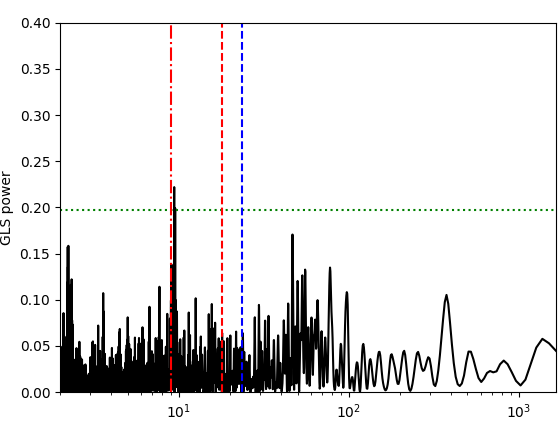}} \\
   \subfloat[][\emph{$V_\text{asy}$}]
   {\includegraphics[width=.39\textwidth]{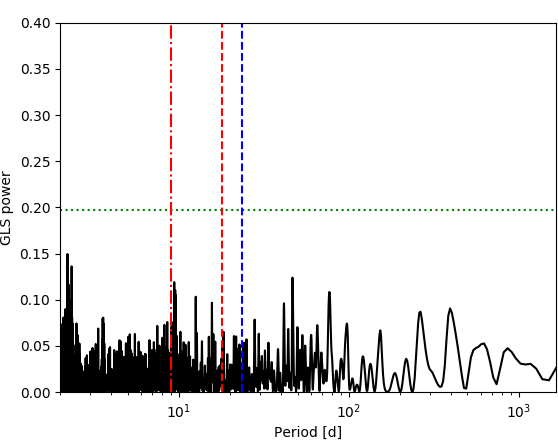}}
      \caption{GLS periodograms of the asymmetry indicators.
      The red vertical lines indicate the potential rotation period (dashed) and its first harmonic (dot-dashed), while the dashed blue vertical lines mark the orbital period of the planetary candidate discussed in Sec. \ref{sec_rv_period}. The dotted horizontal lines indicate the 1$\%$ FAP level.}
         \label{asy_per_plot}
\end{figure}

Then, to expand the analysis of the stellar activity signals of GJ 685 performed by \citet{suarezmascarenoetal2018}, we derived the complete time series for all the line profile indicators evaluated with the method from \citet{lanzaetal2018}, who derived several line profile asymmetry indicators by computing the cross-correlation function (CCF) between a mask and the stellar spectra. For our study, we selected three asymmetry indicators: the bisector inverse span (BIS), $\Delta V$ (which compute the difference between the RV values computed with a Gaussian and bi-Gaussian best fit of the CCF respectively), and $V_\text{asy}$ (which quantifies the asymmetry in the radial-velocity spectral line information content). In addition to this asymmetry analysis, we also derived the activity indexes based on the stellar Ca~{\sc ii}  H \& K, H$\alpha$, Na~{\sc i} D$_{\rm 1}$ D$_{\rm 2}$, and He~{\sc i} D$_{\rm 3}$ spectral lines, following the procedure described in \citet{gomesdasilva11}.

As a first order test of the effect of stellar chromospheric activity on the RV time series, we checked for correlations between the asymmetry and activity indexes, and the RV datasets. We computed the Pearson correlation coefficients for the different combinations of RV and activity indicators, and no significant correlation was identified ($\left | \rho \right | \lesssim 0.3$ for all the indexes).

\begin{figure*}
   \centering
   \subfloat[][\emph{Ca~{\sc ii}  H \& K}]
   {\includegraphics[width=.39\textwidth]{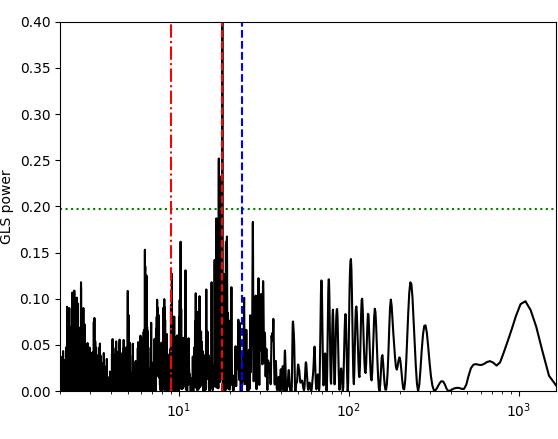}}
   \subfloat[][\emph{H$\alpha$}]
   {\includegraphics[width=.39\textwidth]{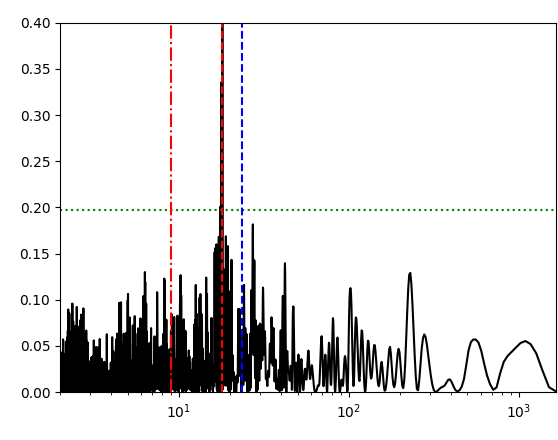}} \\
   \subfloat[][\emph{Na~{\sc i} D$_{\rm 1}$ D$_{\rm 2}$}]
   {\includegraphics[width=.39\textwidth]{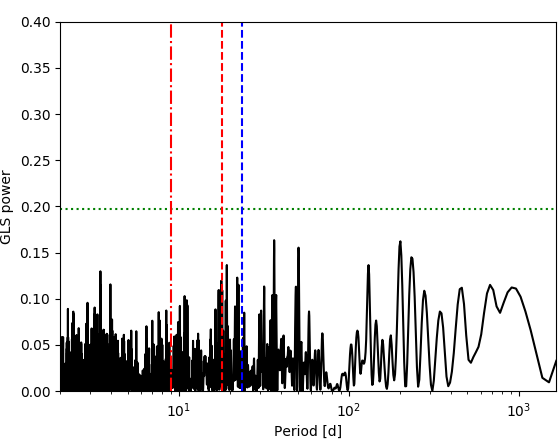}}
   \subfloat[][\emph{He~{\sc i} D$_{\rm 3}$}]
   {\includegraphics[width=.39\textwidth]{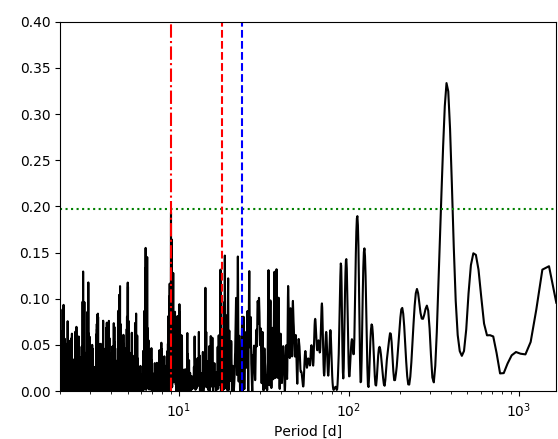}}
      \caption{GLS periodograms of the activity indexes.
      The red vertical lines indicate the potential rotation period (dashed) and its first harmonic (dot-dashed), while the dashed blue vertical lines mark the orbital period of the planetary candidate discussed in Sec. \ref{sec_rv_period}. The dotted horizontal lines indicate the 1$\%$ FAP level.}
         \label{act_per_plot}
\end{figure*}

The GLS periodograms of these asymmetry and activity indexes are shown in Figure \ref{asy_per_plot} and \ref{act_per_plot}. It is worth noticing that, for the asymmetry indicator $V_\text{asy}$ we show the results obtained from the new definition by \citet{lanzaetal2018} ($V_\text{asy(mod)}$), since, as they stated, the original definition from \citet{figueiraetal2013} is sensitive to genuine Doppler shifts of the star, thus presenting misleading signals at the periods of actual planets, leading to erroneous rejections.

We can see in Fig. \ref{asy_per_plot} that the asymmetry indicators show only weak periodic signals, which for two of them, BIS and $\Delta V$, correspond to the first harmonic of the stellar rotation period. In Fig. \ref{act_per_plot} we see instead that the Ca~{\sc ii}  H \& K and H$\alpha$ present strong signals corresponding to the stellar rotation, while the He~{\sc i} D$_{\rm 3}$ index shows only a long term signal around 400 d.\footnote{In the analysis of the residuals of the He~{\sc i} D$_{\rm 3}$ time series (not shown), only an additional signal at $ P = 9$ d emerges, corresponding to the rotation period first harmonic.} The GLS analysis of the Na~{\sc i} D$_{\rm 1}$ D$_{\rm 2}$ time series produces no significant signals.

Our GLS analysis of different asymmetry and activity indicators confirms the results of \citet{suarezmascarenoetal2018}, providing a slightly longer stellar rotation period of 18 d, with clear signals at both the rotation period and its first harmonic. This confirms the rotation period, $P_\text{rot} = 18$ d, identified in the RV time series in Sect. \ref{sec_rv_period}, with respect to the shorter $P_\text{rot, AP} = 16.85$ d found in the APACHE photometric data in Sect. \ref{apache_photo}.  Moreover it is worth noticing how no peak was identified in the indicators time series at periods corresponding or close to the $P_b = 23$ d period of the new planet candidate GJ 685\,b.

\section{MCMC analysis}
\label{mcmc_model}

We then proceded to expand our analyses of the RV and activity indexes time series with a combined fit of the Keplerian and stellar activity signals.
A very common method to model and subtract the stellar activity correlated “noise” from RV time series is the Gaussian Process (GP) regression \citep[e.g.][] {haywoodetal2014,grunblattetal2015,dumusque2017,pinamontietal2018}. This technique has proven to be especially effective when adopting a quasi-periodic covariance function, described by four parameters, called hyper-parameters:
\begin{eqnarray}
\label{eq_gpker}
K(t, t^{\prime}) = h^2\cdot\exp\bigg[-\frac{(t-t^{\prime})^2}{2\lambda^2} - \frac{sin^{2}(\dfrac{\pi(t-t^{\prime})}{\theta})}{2w^2}\bigg] + \nonumber \\
+\, (\sigma^{2}_{\rm data}(t)\,+\,\sigma_{\rm jit}^{2})\cdot\delta_{\rm t, t^{\prime}},
\end{eqnarray}
where $t$ and $t^{\prime}$ indicate two different epochs; $h$ is the amplitude of the correlations; $\theta$ represents the period of the correlated signal (and corresponds to the rotation period of the star in our model); $w$ is the length scale of the periodic component; and $\lambda$ is the correlation decay timescale (which can be related to the decay time of the active regions); $\sigma_{\rm data}(t)$ is the data internal error at time $t$ for each instrument; $\sigma_{\rm jit}$ is the additional uncorrelated 'jitter' term, used in the analysis of the RVs; $\delta_{\rm t, t^{\prime}}$ is the Kronecker delta function.

For a more thorough description of the GP kernel and hyper-parameters see \citet{pinamontietal2018}.

We applied the GP regression as part of an MCMC analysis, performed via the publicly available \texttt{emcee} algorithm \citep{foreman13}, and \texttt{GEORGE} Python library  \citep{ambikasaranetal2015}. We used 150 random walkers to sample the parameter space. The posterior distributions have been derived after applying a burn-in as explained in \citet{eastman13} (and references therein). To evaluate the convergence of the different MCMC analyses we calculated the integrated correlation time for each of the parameters, and stopped the code after a number of steps equal to 150 times the largest autocorrelation times of all the parameters \citep{foreman13}.

\subsection{Activity indexes GP analysis}
\label{activity_analysis}

First we performed a GP emcee fit of the activity index time series, as a reference for the following GP+planetary signal analysis of the RV time series. We show only the H$\alpha$ time series since, as shown in Fig. \ref{act_per_plot} the stellar rotation signal of the star is clearly present in the time series, and some studies also suggest the H$\alpha$ to be the best indicators of the activity of early- and mid-M type stars \citep[e.g.][]{robertsonetal2013}. We nonetheless performed also an emcee analysis of the Ca~{\sc ii}  H \& K time series (not shown), which produced analogous results.

\begin{figure}
   \centering
   \includegraphics[width=.45\textwidth]{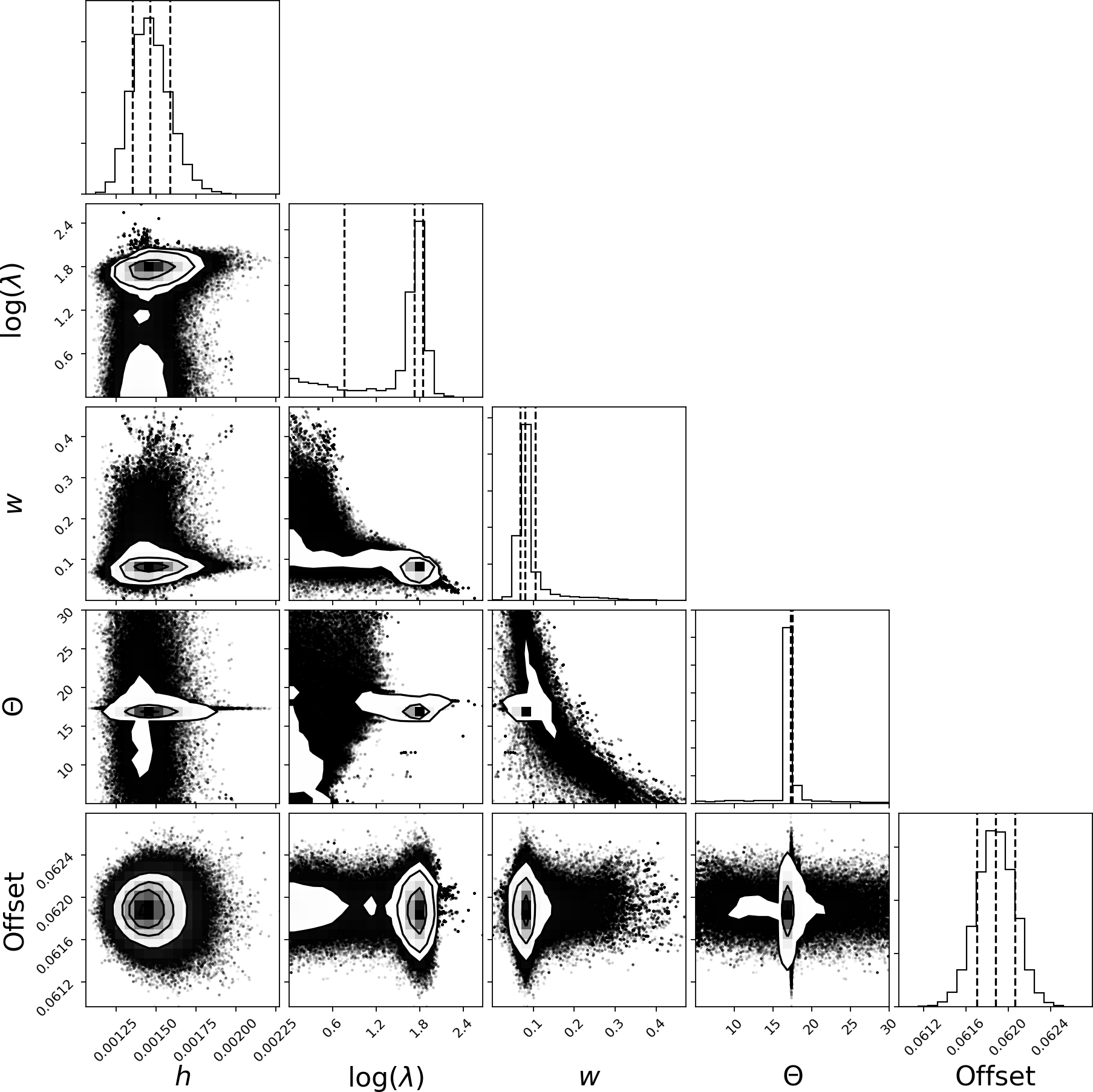}
      \caption{Posterior distributions of the fitted (hyper-)parameters of the GP quasi-periodic model applied to the time series of H$\alpha$ activity index. The
vertical dashed lines denote the median and the 16 th - 84 th percentiles.}
         \label{cahk_gp_plot}
\end{figure}

We chose uniform priors for all the GP hyper-parameters. For the $\lambda$ hyper-parameter, which represents the correlation decay timescale, and can range over several orders of magnitude, we adopted a uniform prior in logarithmic scale, to avoid oversampling the long scales. The adopted priors and best-fit results for the GP hyper-parameters of the analysis are listed in Table \ref{tab-gp-act-ind}, while the a posteriori distributions are shown in Figure \ref{cahk_gp_plot}.

The GP fit confirms the rotation period of the star, corresponding to the hyper-parameter $\theta$, to be close to $\sim 18$ d, as inferred from the GLS analysis. It is also worth noticing that the median value of the $\lambda$ hyper-parameter, which is related to the evolution time-scale of the active regions, is of the order of 3-4 times the rotation period. This is consistent with an independent analysis of a few M dwarfs in the HADES sample reported in \citet{scandariatoetal2017}. Its uncertainties are very large, due to the short time scales tail of the distribution which can be observed in Fig. \ref{cahk_gp_plot}. However, this uncertainty does not affect the determination of the rotation period $\theta$, which is well constrained within a 1 d interval. 
This analysis also does not present any evidence of longer period signals in the activity indexes of GJ 685, in particular around the candidate planet period $P_b = 24$ d. 

\begin{table}
   \caption[]{Priors and best-fit results for the Gaussian process regression analysis of the H$\alpha$ activity indicator.}
          \label{tab-gp-act-ind}
          \centering
         \small
    \begin{tabular}{l l l}
             \hline
             \noalign{\smallskip}
             Jump parameter     &  Prior & Best-fit value  \\
             \hline
             \noalign{\smallskip}
             $h$  & $\mathcal{U}$(0,1.0) & 0.00146$^{+0.00013}_{-0.00011}$ \\
             \noalign{\smallskip}
             $\lambda$ [days] & $\log \mathcal{U}$(1,3\,000) & 53$^{+16}_{-47}$ \\
             \noalign{\smallskip}
             $w$ & $\mathcal{U}$(0,1) & 0.080$^{+0.025}_{-0.011}$ \\
             \noalign{\smallskip}
             $\theta$ [days] & $\mathcal{U}(5,30)$ & 17.34$^{+0.13}_{-0.05}$ \\
             \noalign{\smallskip}
             Offset & $\mathcal{U}(-1.0,3.0)$ & 0.06188$^{+0.00018}_{-0.00018}$ \\
             \noalign{\smallskip}
             \hline
      \end{tabular}
\end{table}

\subsection{emcee analysis of the RV time series}
\label{emcee_analysis}

To better understand the structure of the stellar activity signal in the RV time series we performed a GP analysis of the dataset, adopting very similar priors to those adopted in the analysis of the H$\alpha$ activity index time series (see Table \ref{tab-gp-act-ind}), and also adding an uncorrelated jitter term to the model. In Table \ref{tab_rv_emcee} we can see the chosen priors, as well as the best fit values. Since from the analysis of the H$\alpha$ time series no evidence emerged of long correlation decay timescales, we restricted the prior of $\lambda$ in the interval $[1,500]$ d, again uniform in logarithmic space. Even though the analysis of the activity and asymmetry indexes in the previous sections pointed out the absence of stellar activity signals at periods larger than 20 d and in particular close to the orbital period of the planet candidate, we decided to keep the prior of the rotation period $\theta$ over the interval $[5,30]$ d, as in the analysis described in the previous Section. In this analysis, we add an uncorrelated jitter term $\sigma_\text{jit}$, which takes into account any additional uncorrelated stellar noise that is not corrected by the GP model. This can also be used to take into account possible residual errors from the instrumental drift correction (see Sect. \ref{rv_time_series}) as done in \citet{afferetal2019}.

\begin{figure}
   \centering
   \includegraphics[width=.45\textwidth]{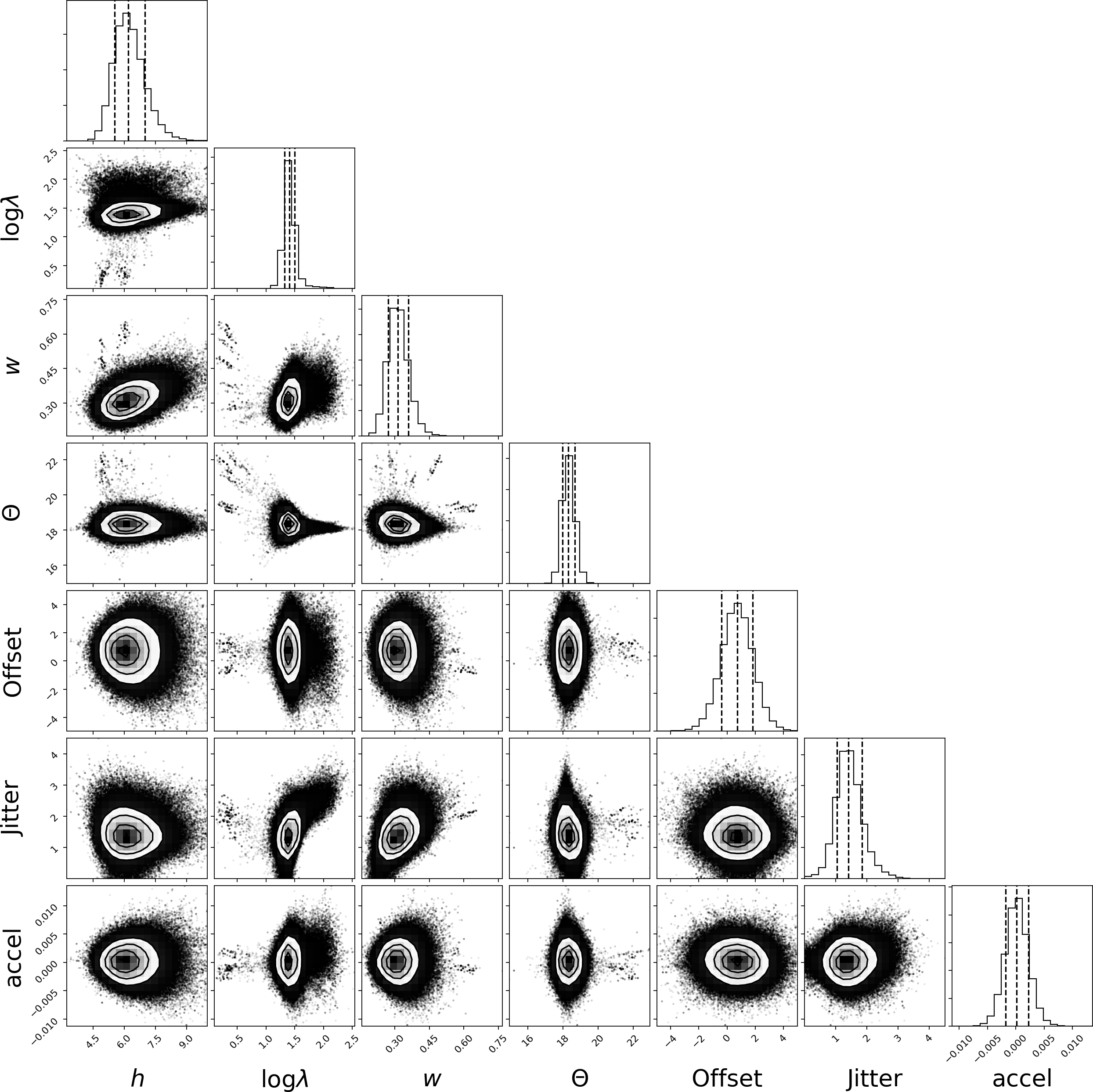}
      \caption{Posterior distributions of the fitted (hyper-)parameters of the pure GP model applied to the RV time series. The
vertical dashed lines denote the median and the 16 th - 84 th percentiles.}
         \label{rv_gp_plot}
\end{figure}

In Fig. \ref{rv_gp_plot} the posterior distributions of the fit parameters are shown. We can see that, even if the GLS periodogram identifies as strongest period $P = 9$ d, the GP correctly identifies the 18 d rotation period of the star, taking into account the complex nature of the stellar activity RV signal due to its quasi-periodic nature. It is also worth noticing in Table \ref{tab_rv_emcee} that the value found in this analysis for the hyper-parameter $\lambda$ is smaller than the one previously found in the analysis of the H$\alpha$ index time series, even if still consistent within the large error bars.

\begin{table}
   \caption[]{Priors and best-fit results for the emcee analysis of GJ 685 RV time series.}
          \label{tab_rv_emcee}
          \centering
         \small
    \begin{tabular}{l l l l}
             \hline
             \noalign{\smallskip}
             Jump parameter     &  Prior & \multicolumn{2}{c}{Best-fit value}  \\
             \noalign{\smallskip}
             & & Pure GP & GP + planet\\
             \hline
             \noalign{\smallskip}
             $h$ [m$/$s]  & $\mathcal{U}$(0,10) & 6.20$^{+0.80}_{-0.66}$ & 6.05$^{+0.94}_{-0.72}$ \\
             \noalign{\smallskip}
             $\lambda$ [days] & $\log \mathcal{U}$(1,500) & 25.4$^{+6.0}_{-4.5}$ & 59$^{+18}_{-14}$ \\
             \noalign{\smallskip}
             $w$ & $\mathcal{U}$(0,1) & 0.312$^{+0.046}_{-0.040}$ & 0.315$^{+0.045}_{-0.041}$ \\
             \noalign{\smallskip}
             $\theta$ [days] & $\mathcal{U}(5,30)$ & 18.30$^{+0.37}_{-0.32}$ & 18.15$^{+0.15}_{-0.16}$ \\
             \noalign{\smallskip}
             Offset [m$/$s] & $\mathcal{U}(-5.0,5.0)$ & 0.7$^{+1.1}_{-1.1}$ & 0.3$^{+1.2}_{-1.3}$ \\
             \noalign{\smallskip}
             Jitter [m$/$s] & $\mathcal{U}(0.0,10.0)$ & 1.41$^{+0.43}_{-0.38}$ & 1.46$^{+0.33}_{-0.32}$ \\
             \noalign{\smallskip}
             Acceleration [m$/$s$\cdot$d] & $\mathcal{U}(-0.05,0.05)$ & 0.0002$^{+0.0020}_{-0.0020}$ & 0.0010$^{+0.0024}_{-0.0024}$ \\
             \noalign{\smallskip}
             \hline
             $k$ [m$/$s] & $\mathcal{U}(0,5.0)$ & - & 3.00$^{+0.53}_{-0.52}$ \\
             \noalign{\smallskip}
             $P$ [days] & $\mathcal{U}(20.0,100.0)$ & - & 24.160$^{+0.061}_{-0.047}$ \\
             \noalign{\smallskip}
             $T_0$ [phase] & $\mathcal{U}(0.0,1.0)$ & - & 0.24$^{+0.11}_{-0.10}$ \\
             \noalign{\smallskip}
             \hline
             Derived parameter & & \\
             \noalign{\smallskip}
             $M_P \sin i$ [M$_\oplus$] & - & - & 9.0$^{+1.7}_{-1.8}$ \\
             \noalign{\smallskip}
             $a_P$ [AU] & - & - & 0.1344$^{+0.0052}_{-0.0051}$ \\
             \noalign{\smallskip}
             \hline
      \end{tabular}
\end{table}

\begin{figure*}
   \centering
   \includegraphics[width=18cm]{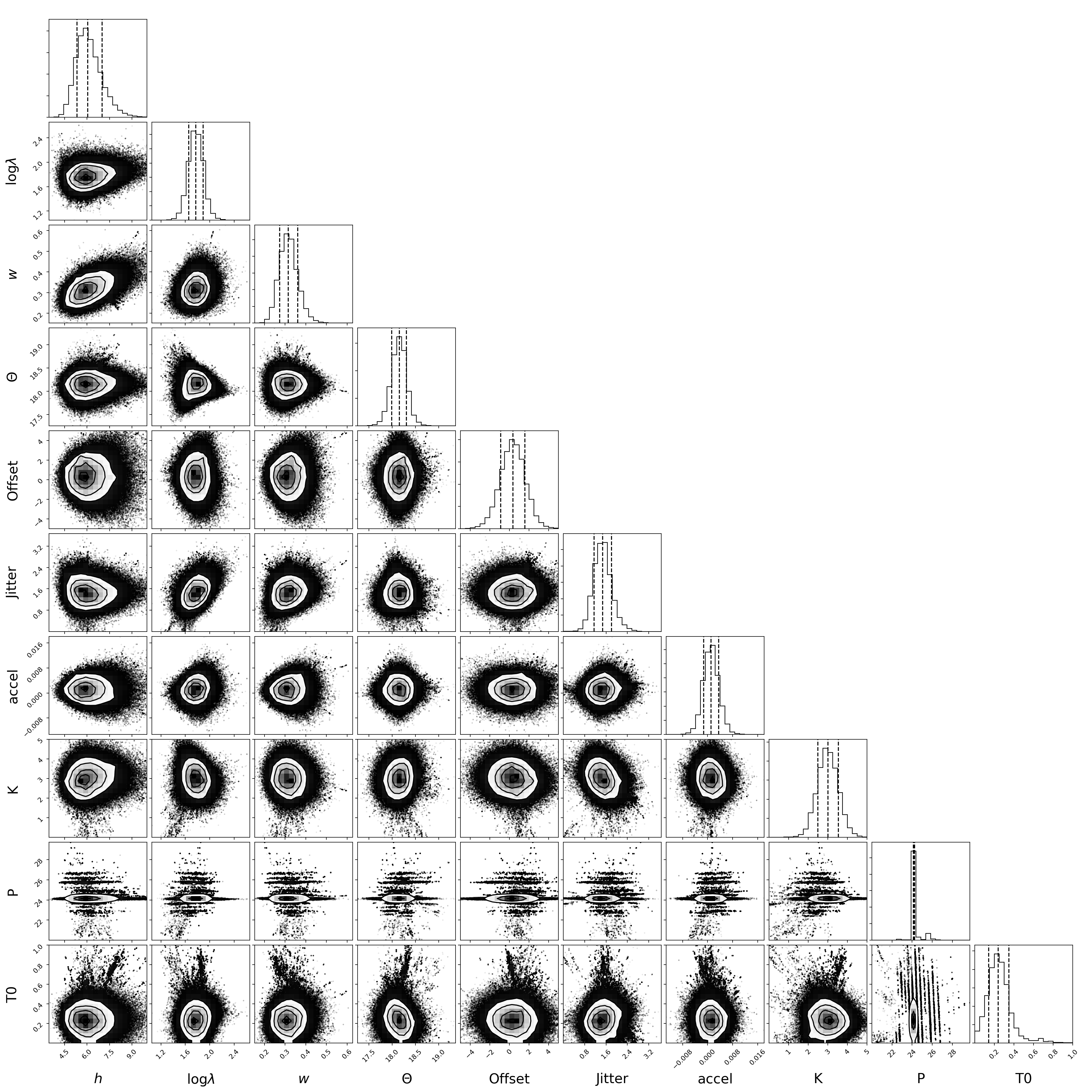}
      \caption{Posterior distributions of the fitted (hyper-)parameters of the GP + 1 planet model applied to the RV time series. The
vertical dashed lines denote the median and the 16 th - 84 th percentiles.}
         \label{rv_gp_oneplan_plot}
\end{figure*}

We then repeated the emcee analysis adding a Keplerian planetary signal to the model, to recover the best fit parameters of the candidate planet GJ 685\,b. We used wide priors for the planetary parameters, not to force the solution on the value found with the GLS periodogram, with the prior of the orbital period ranging up to 100 d. For the same reason, the priors on the planetary parameters were all chosen to be uninformative uniform priors. We can see the best fit solution in the right column of Table \ref{tab_rv_emcee}, while the posterior distributions are shown in Fig. \ref{rv_gp_oneplan_plot}. It is worth noticing that, while the amplitude of the signal is quite similar to the value recovered from the periodogram analysis, the period is slightly longer, due to the simultaneous fitting of the stellar activity signal with the GP. In the bottom right contour plot in Fig. \ref{rv_gp_oneplan_plot} appear to be present a series of aliases of the orbital period $P$, slightly correlated with $T_0$. However, as we can see in the histograms of the posterior distributions of the orbital parameters, these aliases are not significant and do not change the best-fit value of the orbital period. Moreover, we can see in the right column of Table \ref{tab_rv_emcee} that the best-fit value of $\lambda$ is larger than in the pure GP analysis (left column), and much more similar to that found in the analysis of the H$\alpha$: this can be explained with the fact that in the pure-GP model the presence of the un-modelled planetary signal interfere with the fit of the stellar activity, and a shorter decay timescale is needed to compensate this effect; once the planetary signal is taken into account in the analysis, the decay timescale returns to the more accurate value found in tha activity index analysis.
It is also worth noticing that the jitter term retrieved by the emcee analysis is small, $\sigma_\text{jit} = 1.46^{+0.33}_{-0.32}$ m s$^{-1}$, thus suggesting low-levels of uncorrelated stellar noise, as well as confirming the low-levels of residual instrumental drift as discussed in Sect. \ref{rv_time_series}.

\begin{figure}
    \centering
   \includegraphics[width=.45\textwidth]{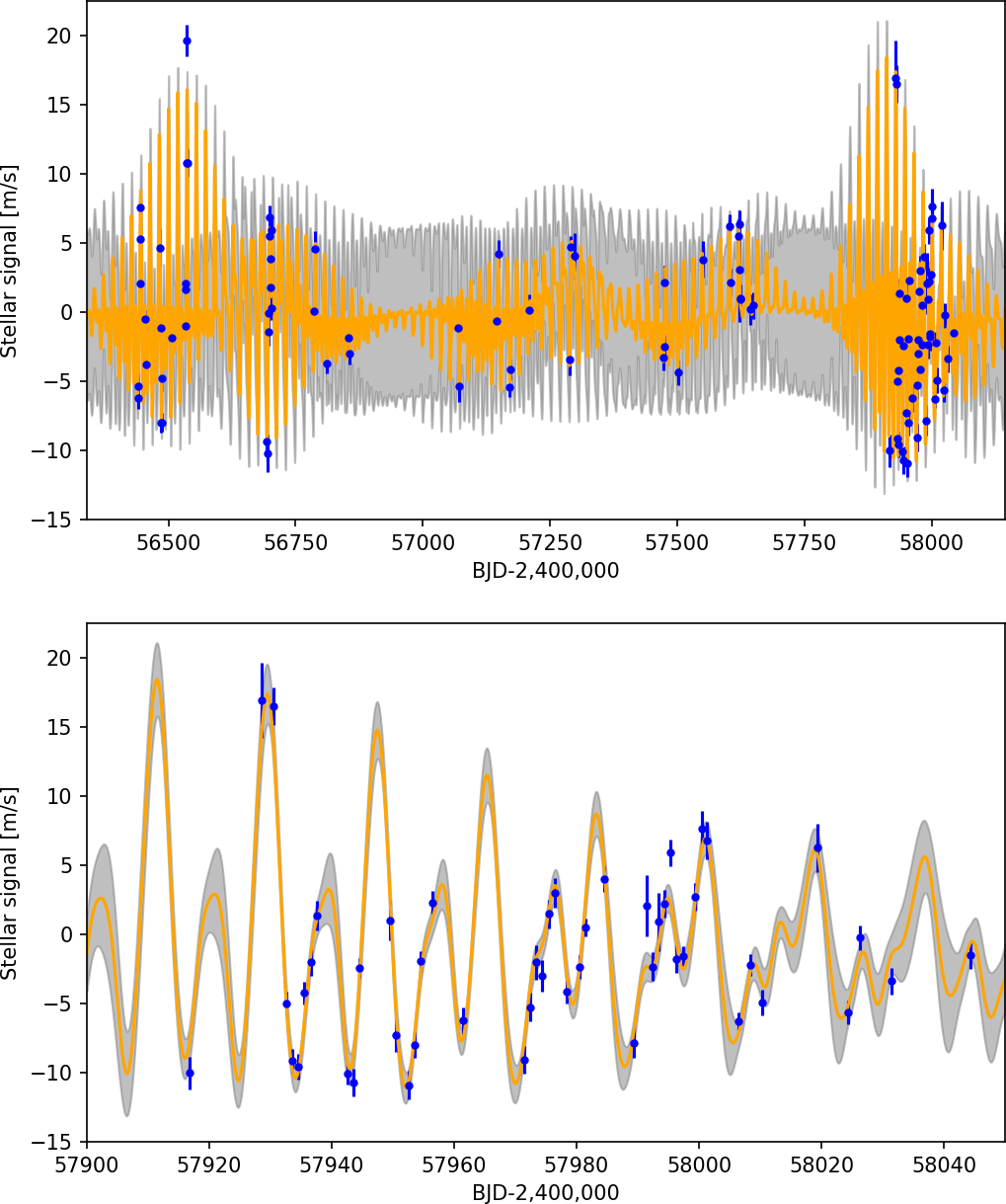}
      \caption{Upper panel: best fit stellar quasi-periodic signal obtained from the GP + 1 planet model (orange line) compared to the RV residuals (blue points). Lower panel: magnification of the last HARPS-N observing season. The grey area represents the $\pm 1 \sigma$ uncertainties of the stellar activity model.}
         \label{gp_model_plot}
\end{figure}

In Fig. \ref{gp_model_plot} is shown the quasi-periodic stellar model, obtained from the simultaneous GP + 1 planet fit, compared to the RV time series residuals after the subtraction of the 24 d planetary signal. We can see both the fine correspondence between the data and our stellar model, and the strength and variability of the stellar signals throughout the four years of HADES observations.

\begin{figure}
    \centering
   \includegraphics[width=.45\textwidth]{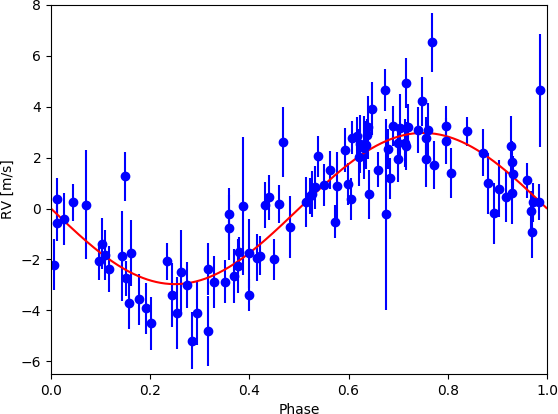}
      \caption{Phase-folded for the RV signal of GJ 685\,b, after the subtraction of the stellar correlated signal.}
         \label{rv_phasefold_plot}
\end{figure}

As additional evidence of the presence of the candidate planet GJ 685\,b we computed the Bayesian Information Criterion \citep[BIC,][]{schwarz1978} for the two models, obtaining BIC $= 635$ and BIC $= 619$ for the pure GP and GP + 1 planet models respectively. There is thus a very strong statistical evidence, $\Delta \text{BIC} = 16$, in favour of the presence of the planetary signal. Fig. \ref{rv_phasefold_plot} shows the phase-folded RV time series, closely following the planetary model after subtraction of the stellar activity signal.

In Table \ref{tab_rv_emcee} are also shown the values of minimum mass, $M_P \sin i$, and semi-major axis, $a_P$, derived from the best-fit orbital parameters of the GP + planet model. The derived minimum mass is $9.0^{+1.7}_{-1.8}$ M$_\oplus$, placing GJ 685\,b within the Super-Earth regime. Since it is known for single-planets systems to show a wide range of eccentricities \citep[e.g.][]{rodigashinz2009,limbachturner2015}, we also tested an eccentric model for GJ 685\,b, in order to constrain the possible eccentricity of the orbit. This additional analysis resulted in a best-fit value of $e = 0.14 ^{+0.18} _{- 0.10}$, consistent with zero within 1.5-$\sigma$, and also with a higher value of BIC $= 628$ with respect to the circular model.

We found no evidence for the presence of additional short-period signals in the RV time series: we computed the GLS periodogram of the RV residuals after the subtraction of the GP + 1 planet model, and no significant signal below the 10$\%$ FAP level was found; additionally, we tested a GP + 2 planets MCMC model, with the second planets parameters free to explore a wide parameter space, and we did not find any dominant signal and no statistical improvement of its BIC over the GP + 1 planet model. Moreover, as reported in Table \ref{rv_gp_oneplan_plot}, the best-fit value of the acceleration in our final model is $0.0010\pm 0.0024$ m s$^{-1}$ d$^{-1}$, largely consistent with zero, thus suggesting the absence of long-period signals.

\section{Summary and discussion}
\label{paperm76_conclusions}

We investigated 106 spectroscopic observations of GJ 685 obtained over 4.4 yr with HARPS-N at the TNG in La Palma, and additional photometry from the APACHE and EXORAP programs. We used RVs derived from the TERRA pipeline, along with activity and asymmetry indexes derived from the same HARPS-N spectra and used to monitor the stellar chromospheric activity of the target.

The radial velocity time series of GJ 685 is dominated by three peaks at 9d, 18d, and 24 d. Our spectroscopic analysis, strengthened by the analyses of two independent photometric light curves of the target, confirm the 18d and 9d period signals to be related to the stellar activity, corresponding respectively to the stellar rotation period and its first harmonic. On the other hand, the 24 d period signal seems not to be related to any stellar effects, and is best described as a Keplerian signal caused by an orbiting planet, GJ 685\,b.

To derive the minimum mass and orbital parameters of GJ 685\,b we fitted the RV time series with a Keplerian model combined with a GP quasi-periodic model to take into account the stellar activity signal. We obtained a period $P_b = 24.160$ d, a semi-major axis $a = 0.1344$ AU, and a minimum mass $M_b \sin i = 9.0$ M$_\oplus$. The GP quasi-periodic model improves the precision of the rotation period of the stars computed by \citet{suarezmascarenoetal2018}, finding a best-fit value of $18.15^{+0.15}_{-0.16}$ d. The amplitude of the stellar activity signal is $h = 6.05^{+0.94}_{-0.72}$ m s$^{-1}$, more than twice the amplitude of the Keplerian signal of GJ 685\,b , similarly to the case of GJ 3942 \citep{pergeretal2017b}.
It is also worth noticing that GJ 685 presents the largest stellar RV signal of the HADES targets with planetary companions detected to date. Moreover, even if the strongest periodic signal present in the RV time series is the rotation period first harmonic, $P_\text{rot$/2$} = 9$ d, the GP model easily identifies the actual rotation period, $P_\text{rot} = 18$ d, of the star as the source of the RV modulation. Even if the prior adopted for the $\theta$ hyper-parameter was very wide, including also the first harmonic value of 9 d, the model converged naturally on the rotation period. This proves once more the effectiveness of GP quasi-periodic models when dealing with complex stellar activity signals producing several different peaks in the periodogram, even if the strongest peak does not corrispond to the stellar rotation period.

We tested an eccentric model for the orbit of GJ 685\,b. It could be expected for a single-planet system to show significant eccentricity, since the orbit of this planet should not be circularized by tides owing to the relatively large separation from the star. The analysis resulted in a best-fit value of eccentricity consistent with zero within 1.5-$\sigma$, and a posterior distribution (not shown) strongly peaked to zero. Moreover, the BIC presented a strong statistical evidence in favour of the circular-orbit model. We thus adopted the null-eccentricity model as the best representation of the orbit of GJ 685\,b.

Due to its close orbit to the host star, GJ 685\,b is unlikely to host an atmosphere capable of maintaining liquid water on its surface. Following the definition of Habitable Zone (HZ) from \citet{kopparapuetal2013}, we computed the inner edge of the HZ for GJ 685, with the most optimistic limits (``recent Venus''), which correspond to a semi-major axis of $a_\text{HZ} = 0.190$ AU, significantly larger than the planet's orbit. 

\begin{figure*}
   \centering
   \includegraphics[width=.45\textwidth]{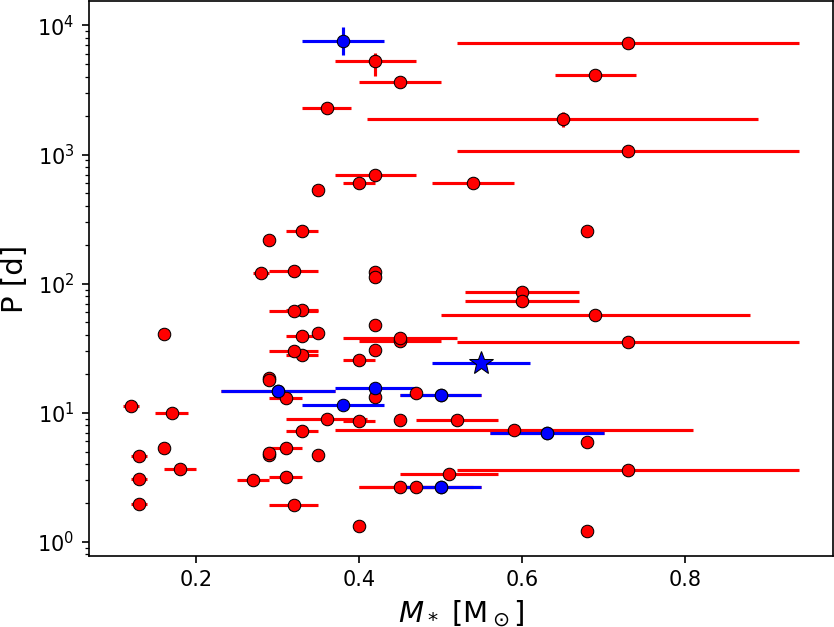}
   \includegraphics[width=.45\textwidth]{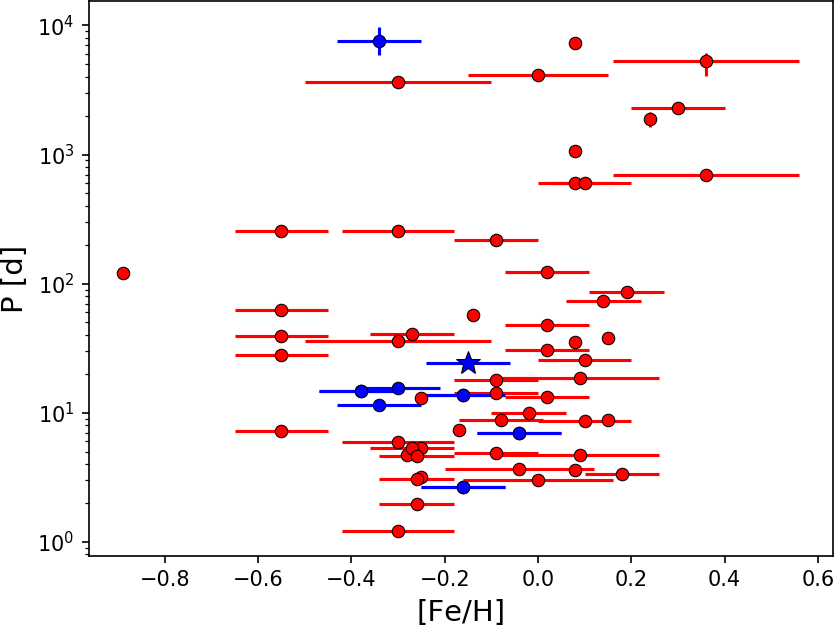} \\
   \includegraphics[width=.45\textwidth]{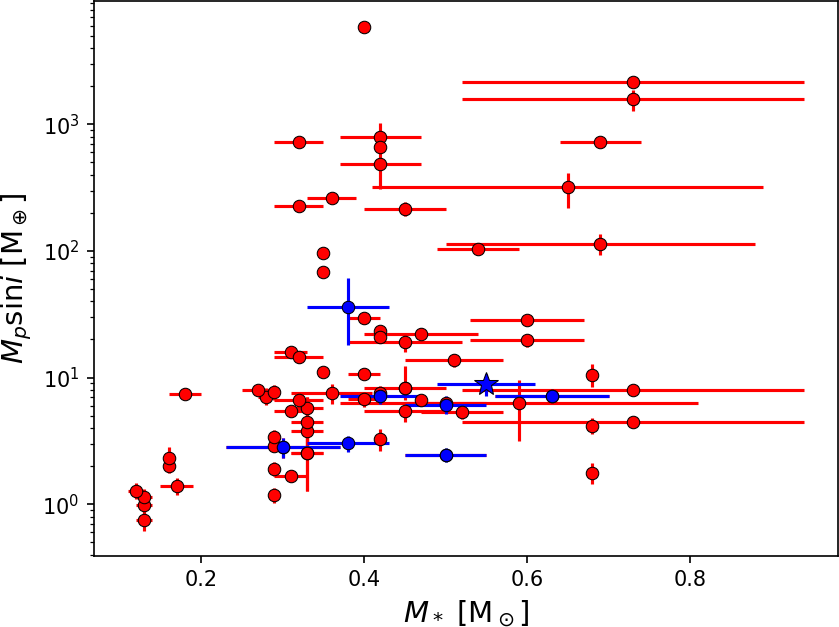}
   \includegraphics[width=.45\textwidth]{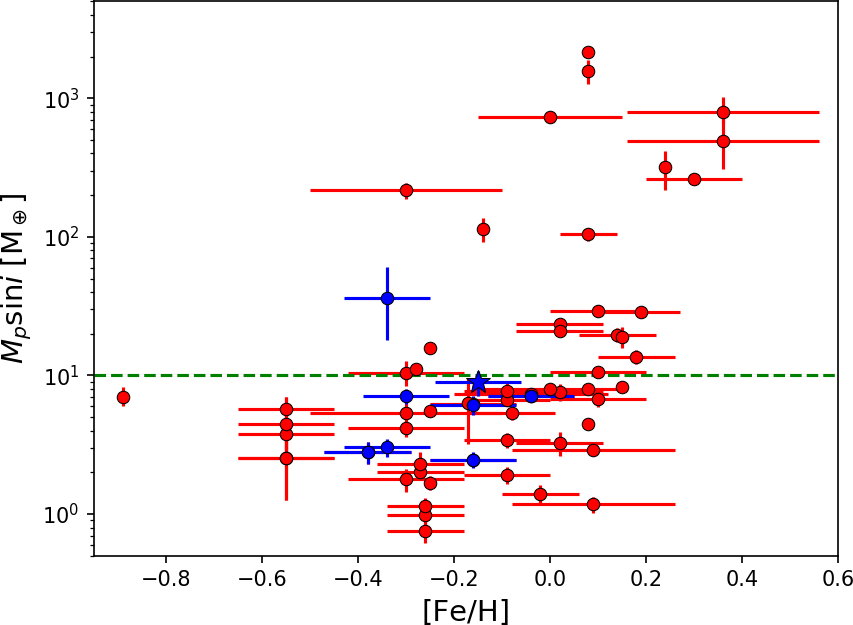}
      \caption{Properties of RV exoplanets orbiting M dwarfs. Top left panel: Orbital period as a function of the mass of the host star; top right panel: orbital period as a function of the metallicity of the host star; bottom left panel: minimum mass as a function of the mass of the host star; bottom right panel: minimum mass as a function of the metallicity of the host star. The red dots represent the M-dwarf hosted RV extrasolar planets, the blue dots represent the HADES planets, and the blue star represents the newly detected planet GJ 685\,b.}
         \label{mdwarfs_planets}
\end{figure*}

\begin{figure}
 \centering
 \includegraphics[width=.4\textwidth]{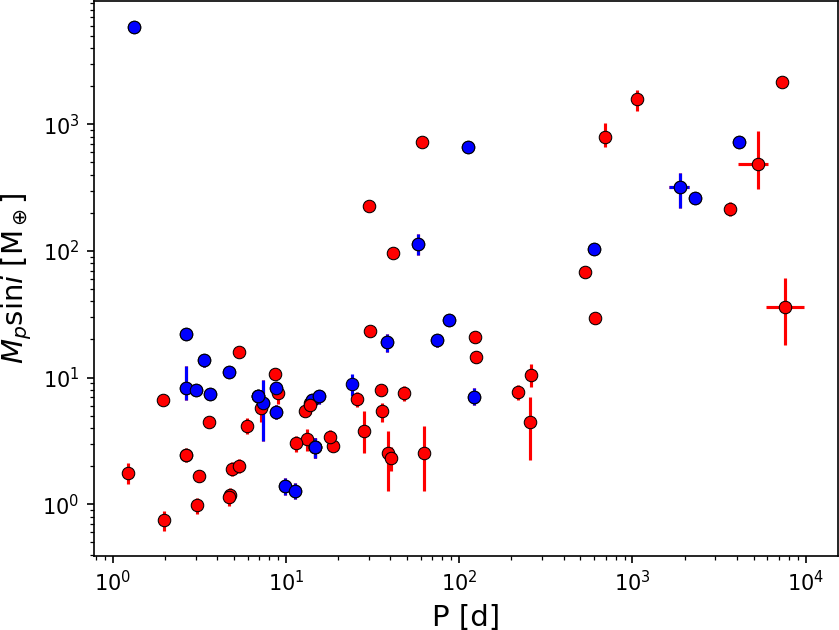}
 \caption{Properties of RV exoplanets orbiting M dwarfs. Minimum mass as a function of the orbital period of the planet. The red dots represent planets orbiting M dwarfs in multiple systems, while the blue dots represent single planets.}
 \label{sing_mult_planets}
\end{figure}

GJ 685\,b is the seventh extrasolar planet discovered within the HADES program \citep{afferetal2016,suarezmascarenoetal2017,pergeretal2017b,pinamontietal2018,afferetal2019}. In Fig. \ref{mdwarfs_planets} and \ref{sing_mult_planets} are shown GJ 685\,b and the other HADES planets compared to the current population of RV detected planets orbiting M dwarfs, selected from the NASA Exoplanet Archive\footnote{\url{https://exoplanetarchive.ipac.caltech.edu/} - 18/12/2018}: the sample is composed of 70 extrasolar planets discovered with the RV method and orbiting stars of spectral type M0 and M9.
Even if the sample is rather small,  some consideration can be made about the planetary and stellar parameters of these systems. First of all, it is worth noticing that there seems to be no high-mass planets orbiting low-mass late-type M dwarfs (bottom left panel Fig. \ref{mdwarfs_planets}). Similarly, there seems to be a scarcity of long period planets orbiting stars with $M_* < 0.3$ M$_\odot$, which is the same threshold below which only low-mass planets have been detected. However, these effects are strongly affected by observational bias, since late-type M dwarfs are very faint and difficult to observe, and thus are usually excluded from RV exoplanet surveys, e.g. the HADES sample has a median stellar mass of 0.5 M$_\odot$, with no target below 0.3 M$_\odot$ \citep{pergeretal2017}. This bias should be solved in the near future, since RV surveys such as CARMENES \citep{quirrenbachetal2014} and SPIRou \citep{moutouetal2017} are intensively monitoring large samples of nearby M dwarfs in search for extrasolar planets, e.g. the CARMENES sample includes $\simeq 120$ stars with $M_* < 0.3$ M$_\odot$ \citep{reinersetal2018}. Nevertheless, to date no such high-mass planet has been announced.

It is also worth noticing, in the bottom-right panel of Fig. \ref{mdwarfs_planets}, that there appears to be a dependence of the planetary minimum mass on the stellar metallicity: we computed the Pearson correlation coefficient, which found a weak-to-moderate correlation, $\rho = 0.30$ $p$-value $= 1.5 \%$. A similar dependence is expected from theoretical models \citep[e.g.][]{mordasinietal2012} and was observed also for solar-mass dwarf stars \citep[e.g.][]{mortieretal2012,wangfischer2015}, while not for planet-hosting giants \citep{mortieretal2013}.
Moreover, this effect was observed to be strong for giant planets, which are much more abundant around high-metallicity stars \citep[e.g.][]{santosetal2005,gaidosmann2014}, but it is still discussed for low-mass Neptune-like planets and super-Earths: some studies found no evidence of correlation between metallicity and planetary occurrence rates for such small planets \citep[e.g.][]{sousaetal2008,gaidosetal2016}, while \citet{wangfischer2015}, analysing a large sample of Kepler-candidates, pointed out that a similar correlation should be present, even if weakened, down to terrestrial planets.
\citet{courcoletal2016} studied a sample of mass-measured exoplanets, and found evidence for the frequency of exo-Neptunes ($M \in [10,40]$ M$_\oplus$) to be correlated with stellar metallicity, while this was not the case for super-Earths ($M < 10$ M$_\oplus$). Similarly, we could divide our sample of RV-detected planets between masses higher and lower than $10$ M$_\oplus$ (green dashed line in the bottom-right panel of Fig. \ref{mdwarfs_planets}): the two sub-samples have both weaker correlations $\rho = 0.23$ $p$-value $= 30 \%$ for $M > 10$ M$_\oplus$ and $\rho = 0.22$ $p$-value $= 14 \%$ for $M < 10$ M$_\oplus$. However, the high-mass subsample contains only 22 planets, which could be the cause of the non-detection of the expected correlation. Thus, based on the selected sample, we are not able to confirm whether the trend observed by \citet{courcoletal2016} affects M dwarfs in the same way as solar-mass stars.

Analysing only the subsample of exoplanets discovered by the HADES program it is difficult to confirm the properties discussed above. The HADES planets are mostly found in a small region of the parameter space, since they usually have low masses and relatively short periods. Also, by construction of the survey, HADES targets are in narrow ranges of stellar parameters \citep{pergeretal2017,maldonadoetal2017}. This will allow a focused analysis of the characteristics of planetary systems around specific stellar hosts, and more results will become available as the analysis of the complete survey's sample draws near. It is worth noticing that the distribution of HADES planets is not surprising, since they are mostly found near the medians of the overall period and minimum mass distributions, $\tilde{P} = 19$ d $\tilde{M} \sin i = 7.6$ M$_\oplus$.

Recently, \citet{luqueetal2018} suggested that masses of single and multiple systems around M dwarfs should follow different distributions. To test this hypothesis, 
in Fig. \ref{sing_mult_planets} are shown the minimum masses and orbital periods of our sample of M-dwarfs planets, distinguishing between single and multiple planetary systems. We can observe that the two populations appear to have similar distributions, with the single planets being on average slightly more massive than those found in multiple systems.\footnote{The minimum mass distribution of detected exoplanets could be affected by observation bias, hiding the presence of additional undetected companions.} To asses the statistical significance of this difference, we performed a two-sided Kolmogorov-Smirnov (K-S) test of the minimum mass distributions of the two populations. We obtained a $p$-value $p = 1.3 \%$, thus reinforcing our hypothesis that planets found in multi- and single-planet systems tend to have different masses. This seems to confirm the effect found by \citet{luqueetal2018}, who studied a different sample of M-dwarf extrasolar planets with masses measured from RVs and TTVs. They suggested two possible explanations, connected to the formation of low-mass planets, i.e. that either \textit{i)} the formation of super-Earth impedes the formation of smaller Earth-like planets in the same system, or \textit{ii)} super-Earth planets around M dwarfs are formed by pile-up of several low-mass planets.
If the former was the case, the single more-massive planet populations should have a mass distribution lower than that of the summed up mass of the planets in multiple systems, due to the absence of additional smaller-mass planets which were not able to form.
We tested this hypothesis on our dataset, performing another K-S test on the two distributions, and obtained a very high $p$-value $p = 47 \%$. We thus found no evidence in favour of the first formation mechanism proposed by \citet{luqueetal2018}, which seems to support the formation of M-dwarf super-Earths by aggregation of smaller-planets.

\begin{figure}
 \centering
 \includegraphics[width=.4\textwidth]{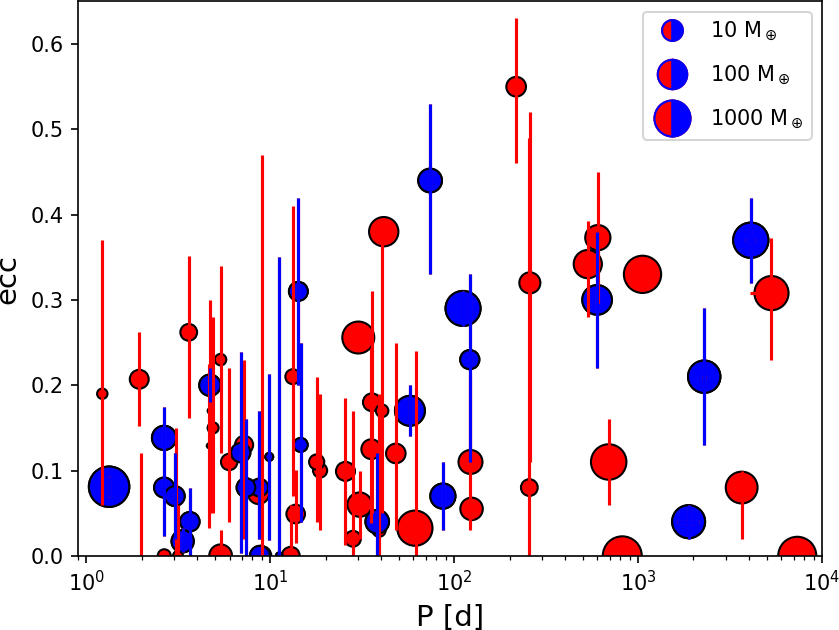}
 \caption{Properties of RV exoplanets orbiting M dwarfs. Orbital eccentricity as a function of the orbital period of the planet. The red dots represent planets orbiting M dwarfs in multiple systems, while the blue dots represent single planets. The marker size represent the minimum mass of the planet.}
 \label{sing_mult_ecc_planets}
\end{figure}

To further compare the properties of the single- and multi-planet populations around M dwarfs, in Fig. \ref{sing_mult_ecc_planets} are shown the eccentricities and orbital periods of the planets in the considered sample. Several studies pointed out a wide distribution of orbital eccentricities of exoplanets, as well as significant correlation with orbital periods \citep[e.g.][]{stepinskiblack2000,kipping2013} and multiplicities \citep[e.g.][]{rodigashinz2009,limbachturner2015}. We can see in Fig. \ref{sing_mult_ecc_planets} that, even if the observed distribution of exoplanets orbiting solar-type stars extend up to $e \sim 0.9$, for M-dwarf RV planets the eccentricities are usually $e < 0.5$, even for long-period giant planets. Moreover, comparing the single- and multi-planet populations, we find no difference in their eeccentricity distribution. They are very similar, with a K-S test suggesting the two distributions to be identical ($p$-value $p = 99 \%$). This is another evidence of the difference between the populations of extrasolar planets orbiting M dwarfs and solar-type stars. Focusing on the sub-sample of low-mass planets ($M_p \sin i < 30$ M$_\oplus$), it is worth noticing that eccentricities are usually poorly constrained: out of 54 low-mass planets, only 7 have eccentricities with a significance higher than 3-$\sigma$: this is not surprising, since eccentricities are often difficult to precisely constrain, in particular for single-planet systems \citep[e.g.][]{wittenmyeretal2013}.

In Fig. \ref{sing_mult_planets} there also appears to be a correlation between the minimum mass and the orbital period, as was suggested by early formation models \citep[and references therein]{zuckermazeh2002}. However, the Pearson correlation coefficient does not favour such correlation in our sample: $\rho = 0.22$ $p$-value $= 5.0 \%$. It is also important to notice that observational biases have a strong influence on the mass and period distributions, since smaller-mass and longer-period planets are more difficult to detect. Recent studies on the occurrence rates of extrasolar planets around M dwarfs, taking into account detection probabilities and detection biases suggest that low-mass long period-planet could be in fact abundant around M dwarfs \citep{bonfils13,tuomi14}, but such an extended analysis is beyond the scope of this paper.
Moreover, it is worth noticing that M stars are a very heterogeneous group, with early- and late-M dwarfs differing both for internal structure and for magnetic activity. Therefore a more accurate study of the properties of planetary systems around small mass stars should consider separately different early- and late-M stars. For this reason, the target of the HADES programme is precisely to study the population of extrasolar planets over a well-defined sample of targets with similar spectral type and stellar properties.

A thorough and unbiased analysis of the detection efficiencies and planetary occurrence rates in the HADES sample, completing and expanding the preliminary statistical analysis from \citet{pergeretal2017}, will be the object of a future publication (Pinamonti et al. in preparation).

\begin{acknowledgements}
GAPS acknowledges support from INAF through the Progetti Premiali funding scheme of the Italian Ministry of Education, University, and Research.
MPi gratefully ackowledges the support from the European Union Seventh Framework Programme (FP7/2007-2013) under Grant Agreement No. 313014 (ETAEARTH).
LA, AFL, GM, JM, and MD acknowledge financial support from Progetto Premiale 2015 FRONTIERA funding scheme of the Italian Ministry of Education, University, and Research. GS acknowledges financial support from ``Accordo ASI-INAF'' No. 2013-016-R.0 July 9, 2013
and July 9, 2015. MPe, and IR, acknowledge support from the Spanish Ministry of Economy and Competitiveness (MINECO) and the Fondo Europeo de Desarrollo Regional (FEDER) through grant ESP2016-80435-C2-1-R, as well as the support of the Generalitat de Catalunya/CERCA program. JIGH, RR and BTP acknowledge financial support from the Spanish Ministry project MINECO AYA2017-86389-P, and JIGH from the Spanish MINECO under the 2013 Ram\'on y Cajal programme MINECO RYC-2013-14875.
This work is based on observations made with the Italian Telescopio Nazionale Galileo (TNG), operated on the island of La Palma by the Fundaci\'on Galielo Galilei of the Istituto Nazionale di Astrofisica (INAF) at the Spanish Observatorio del Roque de los Muchachos (ORM) of the Instituto de Astrofísica de Canarias (IAC).
This work has made use of data from the European Space Agency (ESA) mission {\it Gaia} (\url{https://www.cosmos.esa.int/gaia}), processed by the {\it Gaia} Data Processing and Analysis Consortium (DPAC, \url{https://www.cosmos.esa.int/web/gaia/dpac/consortium}). Funding for the DPAC has been provided by national institutions, in particular the institutions participating in the {\it Gaia} Multilateral Agreement.
We gratefully acknowledge an anonymous referee for her/his insightful comments that materially improved an earlier version of this manuscript.

\end{acknowledgements}

  \bibliographystyle{aa} 
  \bibliography{biblio}

\begin{appendix}

\section{Observation log for GJ 685}

In this Section we report the observational data collected with the HARPS-N spectrograph as part the HADES project and used in the present study. We list in Table \ref{table_rv_data} the observation dates (barycentric Julian date or BJD), the radial velocities (RVs) calculated by the TERRA pipeline \citep{anglada-escudebutler2012}, and asymmetry indexes BIS, $\Delta V$, and $V_\text{asy}$ derived as described in \citet{lanzaetal2018}. The RV errors reported are the formal ones, not including the jitter term.
In Table \ref{table_activity_data} we list the observation dates and the activity indexes Ca~{\sc ii}  H \& K, H$\alpha$, Na~{\sc i} D$_{\rm 1}$ D$_{\rm 2}$, and He~{\sc i} D$_{\rm 3}$, derived following the procedure by \citet{gomesdasilva11}.

\onecolumn
\begin{longtable}{ccccccccc}
\caption{RV and asymmetry indexes data of the 106 observed HARPS-N spectra of GJ 685. We list observation epochs, RVs, BIS, $\Delta V$, $V_\text{asy}$, and the respective errors.}\\
\label{table_rv_data}\\ 
\hline
\hline
BJD$-2400000$ & RV & RV$_\text{Err}$ & BIS & BIS$_\text{Err}$ & $\Delta V$ & $\Delta V_\text{Err}$ & $V_\text{asy}$ & $V_\text{asy, Err}$  \\
$[d]$ & $[$m$/$s$]$ & $[$m$/$s$]$ \\
\hline
\endfirsthead
\caption{Continued.} \\
\hline
\hline
BJD$-2400000$ & RV & RV$_\text{Err}$ & BIS & BIS$_\text{Err}$ & $\Delta V$ & $\Delta V_\text{Err}$ & $V_\text{asy}$ & $V_\text{asy, Err}$  \\
$[d]$ & $[$m$/$s$]$ & $[$m$/$s$]$ \\
\hline
\endhead
\hline
\endfoot
  56439.6165 & $      -3.74$ & $       0.83$ & $     0.0487$ & $     0.0068$ & $      0.060$ & $      0.019$ & $     -0.070$ & $      0.012$ \\ 
  56440.6070 & $      -3.01$ & $       0.89$ & $     0.0423$ & $     0.0069$ & $      0.053$ & $      0.019$ & $     -0.057$ & $      0.012$ \\ 
  56443.4917 & $       3.00$ & $       0.99$ & $     0.0401$ & $     0.0046$ & $      0.050$ & $      0.017$ & $     -0.061$ & $      0.012$ \\ 
  56443.7301 & $       6.01$ & $       1.16$ & $     0.0396$ & $     0.0047$ & $      0.049$ & $      0.016$ & $     -0.061$ & $      0.012$ \\ 
  56444.5211 & $       7.73$ & $       0.68$ & $     0.0416$ & $     0.0056$ & $      0.052$ & $      0.019$ & $     -0.060$ & $      0.012$ \\ 
  56454.6490 & $      -3.00$ & $       0.60$ & $     0.0458$ & $     0.0069$ & $      0.056$ & $      0.018$ & $     -0.065$ & $      0.012$ \\ 
  56455.6660 & $      -5.61$ & $       0.76$ & $     0.0458$ & $     0.0061$ & $      0.058$ & $      0.019$ & $     -0.067$ & $      0.012$ \\ 
  56483.5766 & $       5.62$ & $       1.32$ & $     0.0286$ & $     0.0063$ & $      0.038$ & $      0.020$ & $     -0.038$ & $      0.013$ \\ 
  56484.5451 & $       0.47$ & $       0.75$ & $     0.0465$ & $     0.0063$ & $      0.057$ & $      0.019$ & $     -0.067$ & $      0.012$ \\ 
  56485.5714 & $      -5.93$ & $       0.69$ & $     0.0462$ & $     0.0063$ & $      0.056$ & $      0.018$ & $     -0.068$ & $      0.012$ \\ 
  56486.5469 & $      -5.58$ & $       0.70$ & $     0.0483$ & $     0.0072$ & $      0.059$ & $      0.020$ & $     -0.066$ & $      0.012$ \\ 
  56487.5532 & $      -2.26$ & $       0.90$ & $     0.0409$ & $     0.0063$ & $      0.052$ & $      0.019$ & $     -0.058$ & $      0.012$ \\ 
  56506.5333 & $      -1.73$ & $       0.90$ & $     0.0398$ & $     0.0061$ & $      0.050$ & $      0.020$ & $     -0.051$ & $      0.012$ \\ 
  56533.3947 & $       0.90$ & $       0.64$ & $     0.0412$ & $     0.0067$ & $      0.052$ & $      0.020$ & $     -0.055$ & $      0.012$ \\ 
  56534.4035 & $       4.43$ & $       0.78$ & $     0.0409$ & $     0.0058$ & $      0.052$ & $      0.019$ & $     -0.057$ & $      0.012$ \\ 
  56534.5066 & $       4.02$ & $       0.81$ & $     0.0440$ & $     0.0078$ & $      0.055$ & $      0.019$ & $     -0.061$ & $      0.013$ \\ 
  56535.3775 & $      13.35$ & $       0.91$ & $     0.0434$ & $     0.0063$ & $      0.051$ & $      0.022$ & $     -0.057$ & $      0.012$ \\ 
  56536.5216 & $      22.22$ & $       1.15$ & $     0.0372$ & $     0.0050$ & $      0.046$ & $      0.022$ & $     -0.060$ & $      0.013$ \\ 
  56537.4850 & $      13.14$ & $       0.99$ & $     0.0322$ & $     0.0035$ & $      0.041$ & $      0.018$ & $     -0.051$ & $      0.012$ \\ 
  56693.7377 & $     -12.55$ & $       0.89$ & $     0.0491$ & $     0.0070$ & $      0.064$ & $      0.018$ & $     -0.073$ & $      0.012$ \\ 
  56694.7625 & $     -13.16$ & $       1.37$ & $     0.0456$ & $     0.0068$ & $      0.056$ & $      0.019$ & $     -0.063$ & $      0.013$ \\ 
  56695.7888 & $      -3.93$ & $       1.03$ & $     0.0505$ & $     0.0056$ & $      0.060$ & $      0.020$ & $     -0.080$ & $      0.013$ \\ 
  56696.7350 & $      -2.05$ & $       0.64$ & $     0.0404$ & $     0.0051$ & $      0.052$ & $      0.018$ & $     -0.061$ & $      0.013$ \\ 
  56697.7342 & $       5.58$ & $       0.89$ & $     0.0429$ & $     0.0053$ & $      0.054$ & $      0.019$ & $     -0.064$ & $      0.012$ \\ 
  56698.7363 & $       5.00$ & $       1.22$ & $     0.0379$ & $     0.0055$ & $      0.046$ & $      0.019$ & $     -0.050$ & $      0.012$ \\ 
  56699.6943 & $       4.12$ & $       0.70$ & $     0.0425$ & $     0.0055$ & $      0.052$ & $      0.018$ & $     -0.062$ & $      0.012$ \\ 
  56700.7016 & $       2.75$ & $       0.74$ & $     0.0489$ & $     0.0077$ & $      0.061$ & $      0.021$ & $     -0.066$ & $      0.012$ \\ 
  56701.7040 & $       1.92$ & $       0.83$ & $     0.0504$ & $     0.0075$ & $      0.061$ & $      0.020$ & $     -0.069$ & $      0.012$ \\ 
  56702.7237 & $       8.12$ & $       1.05$ & $     0.0468$ & $     0.0064$ & $      0.060$ & $      0.020$ & $     -0.074$ & $      0.013$ \\ 
  56786.5800 & $      -2.18$ & $       0.90$ & $     0.0476$ & $     0.0082$ & $      0.058$ & $      0.020$ & $     -0.063$ & $      0.014$ \\ 
  56787.6315 & $       1.84$ & $       1.31$ & $     0.0426$ & $     0.0070$ & $      0.056$ & $      0.019$ & $     -0.062$ & $      0.014$ \\ 
  56811.5545 & $      -6.35$ & $       0.69$ & $     0.0423$ & $     0.0071$ & $      0.055$ & $      0.020$ & $     -0.059$ & $      0.013$ \\ 
  56854.5592 & $      -0.73$ & $       0.73$ & $     0.0488$ & $     0.0082$ & $      0.062$ & $      0.020$ & $     -0.067$ & $      0.013$ \\ 
  56855.5368 & $      -2.65$ & $       0.76$ & $     0.0507$ & $     0.0077$ & $      0.064$ & $      0.020$ & $     -0.073$ & $      0.013$ \\ 
  57069.7411 & $       1.54$ & $       0.57$ & $     0.0467$ & $     0.0077$ & $      0.060$ & $      0.020$ & $     -0.067$ & $      0.014$ \\ 
  57070.7726 & $      -3.18$ & $       1.20$ & $     0.0462$ & $     0.0069$ & $      0.055$ & $      0.019$ & $     -0.060$ & $      0.014$ \\ 
  57145.7379 & $      -0.17$ & $       0.70$ & $     0.0429$ & $     0.0082$ & $      0.056$ & $      0.020$ & $     -0.058$ & $      0.014$ \\ 
  57148.6158 & $       2.58$ & $       1.00$ & $     0.0433$ & $     0.0070$ & $      0.053$ & $      0.022$ & $     -0.058$ & $      0.013$ \\ 
  57170.5973 & $      -5.50$ & $       0.72$ & $     0.0499$ & $     0.0061$ & $      0.064$ & $      0.018$ & $     -0.079$ & $      0.013$ \\ 
  57172.6445 & $      -5.64$ & $       0.73$ & $     0.0513$ & $     0.0062$ & $      0.063$ & $      0.018$ & $     -0.078$ & $      0.013$ \\ 
  57209.5337 & $       2.58$ & $       1.14$ & $     0.0437$ & $     0.0062$ & $      0.055$ & $      0.020$ & $     -0.064$ & $      0.014$ \\ 
  57289.3896 & $      -1.81$ & $       1.18$ & $     0.0354$ & $     0.0063$ & $      0.047$ & $      0.018$ & $     -0.048$ & $      0.015$ \\ 
  57291.4060 & $       4.80$ & $       0.82$ & $     0.0405$ & $     0.0056$ & $      0.050$ & $      0.020$ & $     -0.057$ & $      0.013$ \\ 
  57297.4116 & $       1.51$ & $       1.64$ & $     0.0495$ & $     0.0062$ & $      0.058$ & $      0.023$ & $     -0.078$ & $      0.015$ \\ 
  57472.6602 & $      -2.32$ & $       0.99$ & $     0.0573$ & $     0.0076$ & $      0.067$ & $      0.022$ & $     -0.082$ & $      0.014$ \\ 
  57474.6766 & $      -0.08$ & $       0.82$ & $     0.0440$ & $     0.0059$ & $      0.057$ & $      0.021$ & $     -0.066$ & $      0.013$ \\ 
  57475.6565 & $       5.07$ & $       1.25$ & $     0.0465$ & $     0.0087$ & $      0.056$ & $      0.024$ & $     -0.059$ & $      0.014$ \\ 
  57501.6422 & $      -0.77$ & $       0.94$ & $     0.0519$ & $     0.0068$ & $      0.064$ & $      0.019$ & $     -0.079$ & $      0.013$ \\ 
  57549.7017 & $       7.38$ & $       1.33$ & $     0.0484$ & $     0.0063$ & $      0.062$ & $      0.027$ & $     -0.081$ & $      0.014$ \\ 
  57603.4803 & $       8.16$ & $       0.83$ & $     0.0342$ & $     0.0053$ & $      0.045$ & $      0.019$ & $     -0.053$ & $      0.014$ \\ 
  57604.4710 & $       3.38$ & $       1.01$ & $     0.0335$ & $     0.0048$ & $      0.045$ & $      0.019$ & $     -0.053$ & $      0.013$ \\ 
  57620.4417 & $       8.48$ & $       1.23$ & $     0.0450$ & $     0.0068$ & $      0.055$ & $      0.021$ & $     -0.065$ & $      0.014$ \\ 
  57621.4791 & $       6.49$ & $       3.74$ & $     0.0291$ & $     0.0134$ & $      0.041$ & $      0.021$ & $     -0.024$ & $      0.020$ \\ 
  57622.4623 & $      10.05$ & $       0.98$ & $     0.0423$ & $     0.0048$ & $      0.053$ & $      0.018$ & $     -0.066$ & $      0.013$ \\ 
  57623.4452 & $       4.67$ & $       1.11$ & $     0.0508$ & $     0.0078$ & $      0.061$ & $      0.021$ & $     -0.073$ & $      0.013$ \\ 
  57624.4137 & $       4.62$ & $       0.77$ & $     0.0475$ & $     0.0060$ & $      0.056$ & $      0.022$ & $     -0.065$ & $      0.013$ \\ 
  57644.3556 & $       3.10$ & $       1.11$ & $     0.0479$ & $     0.0058$ & $      0.060$ & $      0.019$ & $     -0.076$ & $      0.013$ \\ 
  57650.3812 & $       3.38$ & $       0.93$ & $     0.0441$ & $     0.0051$ & $      0.052$ & $      0.018$ & $     -0.067$ & $      0.013$ \\ 
  57916.6963 & $      -7.16$ & $       1.19$ & $     0.0496$ & $     0.0089$ & $      0.058$ & $      0.020$ & $     -0.065$ & $      0.014$ \\ 
  57928.6057 & $      16.08$ & $       2.71$ & $     0.0330$ & $     0.0099$ & $      0.038$ & $      0.027$ & $     -0.056$ & $      0.017$ \\ 
  57930.5601 & $      17.05$ & $       1.36$ & $     0.0397$ & $     0.0057$ & $      0.046$ & $      0.029$ & $     -0.067$ & $      0.015$ \\ 
  57932.5778 & $      -2.89$ & $       0.81$ & $     0.0477$ & $     0.0073$ & $      0.059$ & $      0.018$ & $     -0.070$ & $      0.014$ \\ 
  57933.5794 & $      -6.32$ & $       0.86$ & $     0.0432$ & $     0.0045$ & $      0.054$ & $      0.018$ & $     -0.067$ & $      0.013$ \\ 
  57934.5802 & $      -6.24$ & $       0.93$ & $     0.0415$ & $     0.0061$ & $      0.054$ & $      0.019$ & $     -0.062$ & $      0.014$ \\ 
  57935.5355 & $      -0.47$ & $       0.81$ & $     0.0481$ & $     0.0067$ & $      0.057$ & $      0.018$ & $     -0.071$ & $      0.014$ \\ 
  57936.5379 & $       2.00$ & $       0.94$ & $     0.0491$ & $     0.0056$ & $      0.057$ & $      0.018$ & $     -0.074$ & $      0.014$ \\ 
  57937.6435 & $       5.44$ & $       1.06$ & $     0.0462$ & $     0.0056$ & $      0.056$ & $      0.018$ & $     -0.070$ & $      0.013$ \\ 
  57942.6500 & $      -8.47$ & $       0.76$ & $     0.0492$ & $     0.0069$ & $      0.058$ & $      0.020$ & $     -0.069$ & $      0.013$ \\ 
  57943.5814 & $      -9.78$ & $       1.00$ & $     0.0461$ & $     0.0050$ & $      0.057$ & $      0.017$ & $     -0.073$ & $      0.013$ \\ 
  57944.4845 & $      -2.17$ & $       0.71$ & $     0.0383$ & $     0.0052$ & $      0.049$ & $      0.018$ & $     -0.060$ & $      0.013$ \\ 
  57949.5628 & $      -0.90$ & $       1.41$ & $     0.0319$ & $     0.0054$ & $      0.044$ & $      0.019$ & $     -0.046$ & $      0.014$ \\ 
  57950.5477 & $      -9.01$ & $       1.24$ & $     0.0538$ & $     0.0055$ & $      0.065$ & $      0.018$ & $     -0.082$ & $      0.014$ \\ 
  57952.5300 & $     -11.84$ & $       1.05$ & $     0.0552$ & $     0.0062$ & $      0.066$ & $      0.019$ & $     -0.083$ & $      0.013$ \\ 
  57953.4719 & $      -8.35$ & $       0.93$ & $     0.0411$ & $     0.0068$ & $      0.053$ & $      0.018$ & $     -0.061$ & $      0.014$ \\ 
  57954.5128 & $      -1.55$ & $       0.73$ & $     0.0455$ & $     0.0068$ & $      0.059$ & $      0.019$ & $     -0.071$ & $      0.014$ \\ 
  57956.4444 & $       4.21$ & $       0.81$ & $     0.0404$ & $     0.0047$ & $      0.051$ & $      0.019$ & $     -0.061$ & $      0.013$ \\ 
  57961.4980 & $      -2.13$ & $       0.97$ & $     0.0520$ & $     0.0066$ & $      0.064$ & $      0.019$ & $     -0.077$ & $      0.013$ \\ 
  57971.3955 & $     -10.51$ & $       1.00$ & $     0.0434$ & $     0.0064$ & $      0.056$ & $      0.018$ & $     -0.064$ & $      0.013$ \\ 
  57972.4675 & $      -7.03$ & $       1.04$ & $     0.0503$ & $     0.0064$ & $      0.060$ & $      0.018$ & $     -0.073$ & $      0.014$ \\ 
  57973.4635 & $      -3.88$ & $       1.26$ & $     0.0495$ & $     0.0077$ & $      0.057$ & $      0.018$ & $     -0.072$ & $      0.015$ \\ 
  57974.4251 & $      -4.79$ & $       1.13$ & $     0.0441$ & $     0.0070$ & $      0.054$ & $      0.019$ & $     -0.064$ & $      0.014$ \\ 
  57975.4988 & $       0.00$ & $       1.03$ & $     0.0408$ & $     0.0053$ & $      0.050$ & $      0.018$ & $     -0.063$ & $      0.014$ \\ 
  57976.5033 & $       1.99$ & $       1.05$ & $     0.0400$ & $     0.0056$ & $      0.052$ & $      0.018$ & $     -0.062$ & $      0.014$ \\ 
  57978.4249 & $      -3.92$ & $       0.80$ & $     0.0458$ & $     0.0058$ & $      0.056$ & $      0.020$ & $     -0.067$ & $      0.013$ \\ 
  57980.4545 & $      -0.57$ & $       0.86$ & $     0.0422$ & $     0.0063$ & $      0.052$ & $      0.021$ & $     -0.060$ & $      0.013$ \\ 
  57981.4380 & $       3.02$ & $       0.64$ & $     0.0429$ & $     0.0064$ & $      0.054$ & $      0.020$ & $     -0.062$ & $      0.013$ \\ 
  57984.4839 & $       8.01$ & $       0.91$ & $     0.0387$ & $     0.0059$ & $      0.046$ & $      0.019$ & $     -0.054$ & $      0.013$ \\ 
  57989.3821 & $      -5.00$ & $       1.12$ & $     0.0352$ & $     0.0046$ & $      0.044$ & $      0.018$ & $     -0.051$ & $      0.014$ \\ 
  57991.3941 & $       3.43$ & $       2.22$ & $     0.0317$ & $     0.0088$ & $      0.042$ & $      0.026$ & $     -0.052$ & $      0.017$ \\ 
  57992.3841 & $      -1.75$ & $       1.03$ & $     0.0418$ & $     0.0051$ & $      0.050$ & $      0.019$ & $     -0.065$ & $      0.014$ \\ 
  57993.4441 & $       0.71$ & $       2.14$ & $     0.0295$ & $     0.0098$ & $      0.038$ & $      0.019$ & $     -0.050$ & $      0.017$ \\ 
  57994.3785 & $       1.43$ & $       0.97$ & $     0.0420$ & $     0.0058$ & $      0.050$ & $      0.020$ & $     -0.064$ & $      0.014$ \\ 
  57995.3651 & $       4.59$ & $       0.97$ & $     0.0392$ & $     0.0043$ & $      0.048$ & $      0.020$ & $     -0.063$ & $      0.014$ \\ 
  57996.3687 & $      -3.46$ & $       1.00$ & $     0.0367$ & $     0.0058$ & $      0.048$ & $      0.018$ & $     -0.056$ & $      0.014$ \\ 
  57997.4147 & $      -3.41$ & $       0.73$ & $     0.0485$ & $     0.0070$ & $      0.061$ & $      0.018$ & $     -0.074$ & $      0.014$ \\ 
  57999.3666 & $       1.15$ & $       1.01$ & $     0.0468$ & $     0.0065$ & $      0.058$ & $      0.018$ & $     -0.071$ & $      0.013$ \\ 
  58000.4202 & $       6.53$ & $       1.27$ & $     0.0455$ & $     0.0072$ & $      0.051$ & $      0.018$ & $     -0.063$ & $      0.014$ \\ 
  58001.3637 & $       6.19$ & $       1.37$ & $     0.0528$ & $     0.0082$ & $      0.058$ & $      0.019$ & $     -0.074$ & $      0.015$ \\ 
  58006.4189 & $      -3.21$ & $       0.64$ & $     0.0438$ & $     0.0060$ & $      0.053$ & $      0.018$ & $     -0.063$ & $      0.014$ \\ 
  58008.4121 & $       1.76$ & $       0.78$ & $     0.0481$ & $     0.0063$ & $      0.060$ & $      0.020$ & $     -0.070$ & $      0.013$ \\ 
  58010.3947 & $      -0.78$ & $       0.95$ & $     0.0468$ & $     0.0067$ & $      0.059$ & $      0.019$ & $     -0.073$ & $      0.014$ \\ 
  58019.3429 & $       5.08$ & $       1.76$ & $     0.0523$ & $     0.0072$ & $      0.063$ & $      0.019$ & $     -0.080$ & $      0.015$ \\ 
  58024.3564 & $      -6.84$ & $       0.84$ & $     0.0434$ & $     0.0055$ & $      0.051$ & $      0.018$ & $     -0.064$ & $      0.013$ \\ 
  58026.3398 & $      -0.24$ & $       0.90$ & $     0.0464$ & $     0.0067$ & $      0.057$ & $      0.019$ & $     -0.067$ & $      0.013$ \\ 
  58031.3905 & $       0.18$ & $       0.96$ & $     0.0407$ & $     0.0055$ & $      0.050$ & $      0.018$ & $     -0.061$ & $      0.013$ \\ 
  58044.3502 & $      -3.02$ & $       1.00$ & $     0.0411$ & $     0.0053$ & $      0.052$ & $      0.018$ & $     -0.062$ & $      0.014$ \\              
\end{longtable}

\begin{longtable}{ccccccccc}
\caption{Activity indexes data of the 106 observed HARPS-N spectra of GJ 685. We list observation epochs, Ca~{\sc ii}  H \& K, H$\alpha$, Na~{\sc i} D$_{\rm 1}$ D$_{\rm 2}$, He~{\sc i} D$_{\rm 3}$, and the respective errors.}\\
\label{table_activity_data}\\ 
\hline
\hline
BJD$-2400000$ & RV & RV$_\text{Err}$ & BIS & BIS$_\text{Err}$ & $\Delta V$ & $\Delta V_\text{Err}$ & $V_\text{asy}$ & $V_\text{asy, Err}$  \\
\hline
\endfirsthead
\caption{Continued.} \\
\hline
\hline
BJD$-2400000$ & Ca~{\sc ii}  H \& K & Ca~{\sc ii}  H \& K$_\text{Err}$ &  H$\alpha$ &  H$\alpha_\text{Err}$ & Na~{\sc i} D$_{\rm 1}$ D$_{\rm 2}$ & Na~{\sc i} D$_{\rm 1}$ D$_\text{ 2, Err}$ & He~{\sc i} D$_{\rm 3}$ & He~{\sc i} D$_\text{ 3, Err}$  \\
\hline
\endhead
\hline
\endfoot
  56439.6165 & $    0.09561$ & $    0.00064$ & $   0.062060$ & $   0.000107$ & $   0.004638$ & $   0.000023$ & $    0.04049$ & $    0.00013$ \\ 
  56440.6070 & $    0.08841$ & $    0.00065$ & $   0.060442$ & $   0.000110$ & $   0.004583$ & $   0.000024$ & $    0.04043$ & $    0.00014$ \\ 
  56443.4917 & $    0.08535$ & $    0.00092$ & $   0.059510$ & $   0.000147$ & $   0.004490$ & $   0.000033$ & $    0.03944$ & $    0.00020$ \\ 
  56443.7301 & $    0.07298$ & $    0.00083$ & $   0.059985$ & $   0.000157$ & $   0.004671$ & $   0.000036$ & $    0.03949$ & $    0.00021$ \\ 
  56444.5211 & $    0.08791$ & $    0.00062$ & $   0.060539$ & $   0.000110$ & $   0.004570$ & $   0.000023$ & $    0.03946$ & $    0.00013$ \\ 
  56454.6490 & $    0.10000$ & $    0.00070$ & $   0.063561$ & $   0.000111$ & $   0.004716$ & $   0.000024$ & $    0.03960$ & $    0.00014$ \\ 
  56455.6660 & $    0.10666$ & $    0.00082$ & $   0.065671$ & $   0.000129$ & $   0.004852$ & $   0.000027$ & $    0.04068$ & $    0.00016$ \\ 
  56483.5766 & $    0.10201$ & $    0.00144$ & $   0.062962$ & $   0.000228$ & $   0.004861$ & $   0.000049$ & $    0.03965$ & $    0.00028$ \\ 
  56484.5451 & $    0.10009$ & $    0.00075$ & $   0.062777$ & $   0.000118$ & $   0.004669$ & $   0.000025$ & $    0.03937$ & $    0.00015$ \\ 
  56485.5714 & $    0.09804$ & $    0.00069$ & $   0.062531$ & $   0.000102$ & $   0.004577$ & $   0.000023$ & $    0.04056$ & $    0.00014$ \\ 
  56486.5469 & $    0.09839$ & $    0.00066$ & $   0.063194$ & $   0.000114$ & $   0.004696$ & $   0.000024$ & $    0.04052$ & $    0.00014$ \\ 
  56487.5532 & $    0.10118$ & $    0.00082$ & $   0.063952$ & $   0.000127$ & $   0.004697$ & $   0.000027$ & $    0.03950$ & $    0.00016$ \\ 
  56506.5333 & $    0.10125$ & $    0.00086$ & $   0.063597$ & $   0.000147$ & $   0.004881$ & $   0.000031$ & $    0.03944$ & $    0.00017$ \\ 
  56533.3947 & $    0.08336$ & $    0.00052$ & $   0.060286$ & $   0.000099$ & $   0.004542$ & $   0.000020$ & $    0.03930$ & $    0.00012$ \\ 
  56534.4035 & $    0.08413$ & $    0.00072$ & $   0.060381$ & $   0.000130$ & $   0.004532$ & $   0.000027$ & $    0.03939$ & $    0.00016$ \\ 
  56534.5066 & $    0.08706$ & $    0.00094$ & $   0.060720$ & $   0.000147$ & $   0.004551$ & $   0.000032$ & $    0.03936$ & $    0.00019$ \\ 
  56535.3775 & $    0.09179$ & $    0.00072$ & $   0.062070$ & $   0.000141$ & $   0.004758$ & $   0.000028$ & $    0.04070$ & $    0.00016$ \\ 
  56536.5216 & $    0.09398$ & $    0.00080$ & $   0.062810$ & $   0.000099$ & $   0.004211$ & $   0.000023$ & $    0.04024$ & $    0.00015$ \\ 
  56537.4850 & $    0.10041$ & $    0.00074$ & $   0.063841$ & $   0.000108$ & $   0.004498$ & $   0.000024$ & $    0.03931$ & $    0.00014$ \\ 
  56693.7377 & $    0.09503$ & $    0.00091$ & $   0.062017$ & $   0.000138$ & $   0.004724$ & $   0.000030$ & $    0.04078$ & $    0.00018$ \\ 
  56694.7625 & $    0.08591$ & $    0.00087$ & $   0.059852$ & $   0.000137$ & $   0.004475$ & $   0.000031$ & $    0.04069$ & $    0.00019$ \\ 
  56695.7888 & $    0.08653$ & $    0.00107$ & $   0.060192$ & $   0.000152$ & $   0.004313$ & $   0.000037$ & $    0.04175$ & $    0.00023$ \\ 
  56696.7350 & $    0.08568$ & $    0.00088$ & $   0.061051$ & $   0.000138$ & $   0.004478$ & $   0.000031$ & $    0.04077$ & $    0.00019$ \\ 
  56697.7342 & $    0.09726$ & $    0.00080$ & $   0.063246$ & $   0.000128$ & $   0.004833$ & $   0.000028$ & $    0.04118$ & $    0.00016$ \\ 
  56698.7363 & $    0.08939$ & $    0.00096$ & $   0.061495$ & $   0.000154$ & $   0.004597$ & $   0.000034$ & $    0.04087$ & $    0.00020$ \\ 
  56699.6943 & $    0.09492$ & $    0.00068$ & $   0.062424$ & $   0.000099$ & $   0.004566$ & $   0.000022$ & $    0.04067$ & $    0.00013$ \\ 
  56700.7016 & $    0.09748$ & $    0.00070$ & $   0.063375$ & $   0.000118$ & $   0.004792$ & $   0.000024$ & $    0.04075$ & $    0.00014$ \\ 
  56701.7040 & $    0.09479$ & $    0.00074$ & $   0.063197$ & $   0.000128$ & $   0.004800$ & $   0.000027$ & $    0.04074$ & $    0.00015$ \\ 
  56702.7237 & $    0.11418$ & $    0.00093$ & $   0.066657$ & $   0.000113$ & $   0.004633$ & $   0.000027$ & $    0.04104$ & $    0.00016$ \\ 
  56786.5800 & $    0.08328$ & $    0.00091$ & $   0.059716$ & $   0.000160$ & $   0.004532$ & $   0.000034$ & $    0.04068$ & $    0.00020$ \\ 
  56787.6315 & $    0.09222$ & $    0.00130$ & $   0.061135$ & $   0.000202$ & $   0.004665$ & $   0.000045$ & $    0.03936$ & $    0.00026$ \\ 
  56811.5545 & $    0.09430$ & $    0.00076$ & $   0.061798$ & $   0.000121$ & $   0.004695$ & $   0.000027$ & $    0.04055$ & $    0.00016$ \\ 
  56854.5592 & $    0.09375$ & $    0.00066$ & $   0.061102$ & $   0.000109$ & $   0.004712$ & $   0.000024$ & $    0.03958$ & $    0.00014$ \\ 
  56855.5368 & $    0.08779$ & $    0.00060$ & $   0.060005$ & $   0.000102$ & $   0.004647$ & $   0.000022$ & $    0.04046$ & $    0.00013$ \\ 
  57069.7411 & $    0.08977$ & $    0.00110$ & $   0.061213$ & $   0.000173$ & $   0.004606$ & $   0.000038$ & $    0.04046$ & $    0.00023$ \\ 
  57070.7726 & $    0.08571$ & $    0.00127$ & $   0.060933$ & $   0.000184$ & $   0.004502$ & $   0.000043$ & $    0.04093$ & $    0.00026$ \\ 
  57145.7379 & $    0.08263$ & $    0.00068$ & $   0.060525$ & $   0.000123$ & $   0.004541$ & $   0.000026$ & $    0.04067$ & $    0.00016$ \\ 
  57148.6158 & $    0.08624$ & $    0.00074$ & $   0.061605$ & $   0.000137$ & $   0.004673$ & $   0.000029$ & $    0.04053$ & $    0.00017$ \\ 
  57170.5973 & $    0.09509$ & $    0.00070$ & $   0.062790$ & $   0.000103$ & $   0.004510$ & $   0.000024$ & $    0.04050$ & $    0.00014$ \\ 
  57172.6445 & $    0.09365$ & $    0.00083$ & $   0.061914$ & $   0.000115$ & $   0.004510$ & $   0.000027$ & $    0.04051$ & $    0.00016$ \\ 
  57209.5337 & $    0.09108$ & $    0.00105$ & $   0.062632$ & $   0.000163$ & $   0.004586$ & $   0.000036$ & $    0.03964$ & $    0.00021$ \\ 
  57289.3896 & $    0.07381$ & $    0.00106$ & $   0.058912$ & $   0.000189$ & $   0.004641$ & $   0.000044$ & $    0.03942$ & $    0.00025$ \\ 
  57291.4060 & $    0.08401$ & $    0.00064$ & $   0.061866$ & $   0.000114$ & $   0.004513$ & $   0.000024$ & $    0.03951$ & $    0.00014$ \\ 
  57297.4116 & $    0.09814$ & $    0.00156$ & $   0.061781$ & $   0.000172$ & $   0.004439$ & $   0.000046$ & $    0.03952$ & $    0.00029$ \\ 
  57472.6602 & $    0.08774$ & $    0.00103$ & $   0.060410$ & $   0.000181$ & $   0.004625$ & $   0.000040$ & $    0.04072$ & $    0.00023$ \\ 
  57474.6766 & $    0.08848$ & $    0.00072$ & $   0.061243$ & $   0.000131$ & $   0.004573$ & $   0.000028$ & $    0.04046$ & $    0.00016$ \\ 
  57475.6565 & $    0.08742$ & $    0.00101$ & $   0.060990$ & $   0.000194$ & $   0.004677$ & $   0.000042$ & $    0.04057$ & $    0.00024$ \\ 
  57501.6422 & $    0.10231$ & $    0.00121$ & $   0.063452$ & $   0.000173$ & $   0.004777$ & $   0.000040$ & $    0.04057$ & $    0.00023$ \\ 
  57549.7017 & $    0.09461$ & $    0.00118$ & $   0.061545$ & $   0.000127$ & $   0.004290$ & $   0.000035$ & $    0.03937$ & $    0.00022$ \\ 
  57603.4803 & $    0.10459$ & $    0.00082$ & $   0.063254$ & $   0.000107$ & $   0.004632$ & $   0.000026$ & $    0.03967$ & $    0.00016$ \\ 
  57604.4710 & $    0.09771$ & $    0.00114$ & $   0.062192$ & $   0.000169$ & $   0.004651$ & $   0.000039$ & $    0.03966$ & $    0.00023$ \\ 
  57620.4417 & $    0.09116$ & $    0.00097$ & $   0.061652$ & $   0.000164$ & $   0.004689$ & $   0.000037$ & $    0.03947$ & $    0.00021$ \\ 
  57621.4791 & $    0.08421$ & $    0.00227$ & $   0.062241$ & $   0.000379$ & $   0.004968$ & $   0.000094$ & $    0.04003$ & $    0.00054$ \\ 
  57622.4623 & $    0.09905$ & $    0.00096$ & $   0.063066$ & $   0.000138$ & $   0.004692$ & $   0.000033$ & $    0.03963$ & $    0.00019$ \\ 
  57623.4452 & $    0.09386$ & $    0.00090$ & $   0.061852$ & $   0.000149$ & $   0.004709$ & $   0.000033$ & $    0.03943$ & $    0.00019$ \\ 
  57624.4137 & $    0.09613$ & $    0.00072$ & $   0.062649$ & $   0.000132$ & $   0.004748$ & $   0.000028$ & $    0.03951$ & $    0.00016$ \\ 
  57644.3556 & $    0.10144$ & $    0.00088$ & $   0.064186$ & $   0.000127$ & $   0.004679$ & $   0.000030$ & $    0.03938$ & $    0.00017$ \\ 
  57650.3812 & $    0.09179$ & $    0.00088$ & $   0.061558$ & $   0.000141$ & $   0.004621$ & $   0.000032$ & $    0.03931$ & $    0.00019$ \\ 
  57916.6963 & $    0.08727$ & $    0.00108$ & $   0.061088$ & $   0.000189$ & $   0.004673$ & $   0.000042$ & $    0.04070$ & $    0.00025$ \\ 
  57928.6057 & $    0.09166$ & $    0.00203$ & $   0.061891$ & $   0.000247$ & $   0.004639$ & $   0.000071$ & $    0.03970$ & $    0.00044$ \\ 
  57930.5601 & $    0.10249$ & $    0.00145$ & $   0.063463$ & $   0.000152$ & $   0.004185$ & $   0.000042$ & $    0.03944$ & $    0.00027$ \\ 
  57932.5778 & $    0.10408$ & $    0.00086$ & $   0.064630$ & $   0.000124$ & $   0.004626$ & $   0.000029$ & $    0.04090$ & $    0.00017$ \\ 
  57933.5794 & $    0.09838$ & $    0.00090$ & $   0.062950$ & $   0.000140$ & $   0.004594$ & $   0.000032$ & $    0.04050$ & $    0.00019$ \\ 
  57934.5802 & $    0.09272$ & $    0.00083$ & $   0.061708$ & $   0.000126$ & $   0.004505$ & $   0.000029$ & $    0.04068$ & $    0.00018$ \\ 
  57935.5355 & $    0.10477$ & $    0.00086$ & $   0.065083$ & $   0.000131$ & $   0.004722$ & $   0.000030$ & $    0.04005$ & $    0.00017$ \\ 
  57936.5379 & $    0.09064$ & $    0.00106$ & $   0.061315$ & $   0.000170$ & $   0.004571$ & $   0.000039$ & $    0.03958$ & $    0.00023$ \\ 
  57937.6435 & $    0.08851$ & $    0.00086$ & $   0.060725$ & $   0.000130$ & $   0.004482$ & $   0.000030$ & $    0.03937$ & $    0.00018$ \\ 
  57942.6500 & $    0.08518$ & $    0.00069$ & $   0.059722$ & $   0.000123$ & $   0.004565$ & $   0.000027$ & $    0.04048$ & $    0.00016$ \\ 
  57943.5814 & $    0.10073$ & $    0.00117$ & $   0.063162$ & $   0.000164$ & $   0.004672$ & $   0.000039$ & $    0.04089$ & $    0.00023$ \\ 
  57944.4845 & $    0.08810$ & $    0.00071$ & $   0.060404$ & $   0.000106$ & $   0.004416$ & $   0.000025$ & $    0.04054$ & $    0.00016$ \\ 
  57949.5628 & $    0.09540$ & $    0.00125$ & $   0.062328$ & $   0.000190$ & $   0.004593$ & $   0.000044$ & $    0.04051$ & $    0.00026$ \\ 
  57950.5477 & $    0.09760$ & $    0.00119$ & $   0.063236$ & $   0.000182$ & $   0.004748$ & $   0.000042$ & $    0.04075$ & $    0.00025$ \\ 
  57952.5300 & $    0.09161$ & $    0.00088$ & $   0.061863$ & $   0.000143$ & $   0.004535$ & $   0.000032$ & $    0.04044$ & $    0.00019$ \\ 
  57953.4719 & $    0.09659$ & $    0.00072$ & $   0.062843$ & $   0.000108$ & $   0.004568$ & $   0.000025$ & $    0.04081$ & $    0.00015$ \\ 
  57954.5128 & $    0.08800$ & $    0.00072$ & $   0.060700$ & $   0.000101$ & $   0.004338$ & $   0.000024$ & $    0.03958$ & $    0.00015$ \\ 
  57956.4444 & $    0.08439$ & $    0.00059$ & $   0.059468$ & $   0.000104$ & $   0.004498$ & $   0.000023$ & $    0.03920$ & $    0.00014$ \\ 
  57961.4980 & $    0.08618$ & $    0.00082$ & $   0.060867$ & $   0.000146$ & $   0.004370$ & $   0.000031$ & $    0.04045$ & $    0.00019$ \\ 
  57971.3955 & $    0.09186$ & $    0.00083$ & $   0.061743$ & $   0.000131$ & $   0.004540$ & $   0.000030$ & $    0.04079$ & $    0.00018$ \\ 
  57972.4675 & $    0.09107$ & $    0.00102$ & $   0.062423$ & $   0.000167$ & $   0.004606$ & $   0.000038$ & $    0.04055$ & $    0.00022$ \\ 
  57973.4635 & $    0.08250$ & $    0.00121$ & $   0.061649$ & $   0.000213$ & $   0.004824$ & $   0.000050$ & $    0.04043$ & $    0.00029$ \\ 
  57974.4251 & $    0.08740$ & $    0.00100$ & $   0.061335$ & $   0.000171$ & $   0.004515$ & $   0.000038$ & $    0.04015$ & $    0.00023$ \\ 
  57975.4988 & $    0.08835$ & $    0.00093$ & $   0.060880$ & $   0.000141$ & $   0.004418$ & $   0.000033$ & $    0.03929$ & $    0.00020$ \\ 
  57976.5033 & $    0.08611$ & $    0.00096$ & $   0.060881$ & $   0.000148$ & $   0.004467$ & $   0.000035$ & $    0.03961$ & $    0.00021$ \\ 
  57978.4249 & $    0.08847$ & $    0.00068$ & $   0.061783$ & $   0.000125$ & $   0.004615$ & $   0.000027$ & $    0.04068$ & $    0.00016$ \\ 
  57980.4545 & $    0.08628$ & $    0.00063$ & $   0.060469$ & $   0.000121$ & $   0.004645$ & $   0.000026$ & $    0.04056$ & $    0.00015$ \\ 
  57981.4380 & $    0.08487$ & $    0.00077$ & $   0.060199$ & $   0.000138$ & $   0.004595$ & $   0.000030$ & $    0.03925$ & $    0.00018$ \\ 
  57984.4839 & $    0.10293$ & $    0.00085$ & $   0.064985$ & $   0.000137$ & $   0.004723$ & $   0.000030$ & $    0.03975$ & $    0.00018$ \\ 
  57989.3821 & $    0.09273$ & $    0.00101$ & $   0.062176$ & $   0.000165$ & $   0.004556$ & $   0.000037$ & $    0.04026$ & $    0.00022$ \\ 
  57991.3941 & $    0.09078$ & $    0.00185$ & $   0.061146$ & $   0.000221$ & $   0.004411$ & $   0.000063$ & $    0.04009$ & $    0.00040$ \\ 
  57992.3841 & $    0.08631$ & $    0.00087$ & $   0.060948$ & $   0.000122$ & $   0.004271$ & $   0.000030$ & $    0.03944$ & $    0.00019$ \\ 
  57993.4441 & $    0.08007$ & $    0.00175$ & $   0.061598$ & $   0.000273$ & $   0.005230$ & $   0.000072$ & $    0.03904$ & $    0.00040$ \\ 
  57994.3785 & $    0.08639$ & $    0.00077$ & $   0.060204$ & $   0.000108$ & $   0.004301$ & $   0.000027$ & $    0.03949$ & $    0.00017$ \\ 
  57995.3651 & $    0.09490$ & $    0.00092$ & $   0.062171$ & $   0.000118$ & $   0.004440$ & $   0.000030$ & $    0.03983$ & $    0.00019$ \\ 
  57996.3687 & $    0.08769$ & $    0.00096$ & $   0.060458$ & $   0.000142$ & $   0.004428$ & $   0.000034$ & $    0.03922$ & $    0.00021$ \\ 
  57997.4147 & $    0.08839$ & $    0.00088$ & $   0.060663$ & $   0.000126$ & $   0.004360$ & $   0.000030$ & $    0.04072$ & $    0.00019$ \\ 
  57999.3666 & $    0.08730$ & $    0.00090$ & $   0.060661$ & $   0.000151$ & $   0.004488$ & $   0.000034$ & $    0.03952$ & $    0.00020$ \\ 
  58000.4202 & $    0.09153$ & $    0.00115$ & $   0.061780$ & $   0.000176$ & $   0.004515$ & $   0.000041$ & $    0.03945$ & $    0.00024$ \\ 
  58001.3637 & $    0.09534$ & $    0.00161$ & $   0.061638$ & $   0.000243$ & $   0.004416$ & $   0.000055$ & $    0.03933$ & $    0.00033$ \\ 
  58006.4189 & $    0.09205$ & $    0.00077$ & $   0.062493$ & $   0.000121$ & $   0.004468$ & $   0.000028$ & $    0.04040$ & $    0.00017$ \\ 
  58008.4121 & $    0.09355$ & $    0.00062$ & $   0.062458$ & $   0.000111$ & $   0.004563$ & $   0.000024$ & $    0.03939$ & $    0.00014$ \\ 
  58010.3947 & $    0.09050$ & $    0.00092$ & $   0.061448$ & $   0.000153$ & $   0.004641$ & $   0.000035$ & $    0.04082$ & $    0.00021$ \\ 
  58019.3429 & $    0.09684$ & $    0.00159$ & $   0.064748$ & $   0.000237$ & $   0.004888$ & $   0.000057$ & $    0.03995$ & $    0.00033$ \\ 
  58024.3564 & $    0.09561$ & $    0.00066$ & $   0.062890$ & $   0.000106$ & $   0.004543$ & $   0.000024$ & $    0.04071$ & $    0.00014$ \\ 
  58026.3398 & $    0.09090$ & $    0.00071$ & $   0.061864$ & $   0.000127$ & $   0.004599$ & $   0.000028$ & $    0.03958$ & $    0.00016$ \\ 
  58031.3905 & $    0.08632$ & $    0.00077$ & $   0.061013$ & $   0.000126$ & $   0.004507$ & $   0.000029$ & $    0.03947$ & $    0.00017$ \\ 
  58044.3502 & $    0.08742$ & $    0.00077$ & $   0.061133$ & $   0.000129$ & $   0.004492$ & $   0.000029$ & $    0.04058$ & $    0.00017$ \\               
\end{longtable}

\twocolumn

\end{appendix}

\end{document}